\documentclass[twocolumn]{aastex631}

\usepackage{academicons}
\usepackage{color}
\definecolor{orcidlogocol}{HTML}{A6CE39}

\usepackage{inputenc}
\inputencoding{utf8}
\DeclareUnicodeCharacter{02BC}{\textquotesingle}
\usepackage{hyperref}
\usepackage{amsmath}

\accepted{April 3, 2024}

\begin{document}
\title{The Impact-driven Atmospheric Loss of Super-Earths \\
around Different Spectral Type Host Stars}

\correspondingauthor{Cong Yu }
\email{yucong@mail.sysu.edu.cn}
\correspondingauthor{Shi Jia }
\email{sjia@must.edu.mo}

\author[0000-0002-0447-7207]{Wei Zhong}
\affiliation{School of Physics and Astronomy, Sun Yat-Sen University, Zhuhai, 519082, People's Republic of China}
\affiliation{CSST Science Center for the Guangdong-Hong Kong-Macau Greater Bay Area, Zhuhai, 519082, People's Republic of China}
\affiliation{State Key Laboratory of Lunar and Planetary Sciences, Macau University of Science and Technology, Macau, People's Republic of China}

\author[0000-0003-0454-7890]{Cong Yu }
\affiliation{School of Physics and Astronomy, Sun Yat-Sen University, Zhuhai, 519082, People's Republic of China}
\affiliation{CSST Science Center for the Guangdong-Hong Kong-Macau Greater Bay Area, Zhuhai, 519082, People's Republic of China}
\affiliation{State Key Laboratory of Lunar and Planetary Sciences, Macau University of Science and Technology, Macau, People's Republic of China}

\author[0000-0001-6808-638X]{Shi Jia }
\affiliation{State Key Laboratory of Lunar and Planetary Sciences, Macau University of Science and Technology, Macau, People's Republic of China}
\affiliation{CNSA Macau Center for Space Exploration and Science, Macau, Peopleʼs Republic of China}

\author[0000-0002-9442-137X]{Shang-Fei Liu }
\affiliation{School of Physics and Astronomy, Sun Yat-Sen University, Zhuhai, 519082, People's Republic of China}
\affiliation{CSST Science Center for the Guangdong-Hong Kong-Macau Greater Bay Area, Zhuhai, 519082, People's Republic of China}

\begin{abstract}

The planet's mass loss is important for the planet's formation and evolution.  The radius valley (RV) is believed to be triggered by evaporation-induced mass loss.  As an alternative mechanism for the RV,  
the mass loss of post-impact planets is thoroughly investigated in this work. The impact energy is converted to the planet's internal energy, enhancing its core energy and accelerating mass loss and orbital migration. 
As the host star changes from K-type to F-type, the planet's mass loss and orbital migration increase. When the initial gas-to-core mass ratio (GCR) is small, the migration efficiency for planets around K-type stars will increase, which helps to suppress mass loss and retain the planet's mass and radius within a specific range. On the contrary, planets around more massive F-type stars experience more substantial mass loss, potentially leading to complete mass loss, and migrate to orbits with longer periods. Our calculation shows that planets around different spectral types of host stars give rise to an RV ranging from 1.3-2.0 $R_{\oplus}$, consistent with the observed range of 1.3-2.6 $R_{\oplus}$. 
Despite the presence of uncertain parameters,  the planetesimal impact can promote the RV establishment for planets around host stars of different spectral types. 

\end{abstract}


\section{Introduction} 
\label{sec:intro}

Nowadays, over 5000 exoplanets have been discovered \citep{2022MNRAS.Gupta}. 
Short-period super-Earths and sub-Neptunes dominate the population of known exoplanets, with approximately $\sim$ 30\%–50\% of Sun-like stars hosting at least one of these small planets
\citep{2018AJ..Petigura, 2018ApJ...Zhu,2022ApJ..Lee}. 
These super-Earths and mini-Neptunes are situated within 0.05-0.3 astronomical units (AU) from their host stars, possess radii ranging from 1.2-4 times the radius of Earth ($R_{\oplus}$), and are known to have the orbital period less than 100 days \citep{2013ApJS..Batalha,2013ApJ..Petigura}. 
The precise spectroscopic follow-up conducted by the California-Kepler Survey  (CKS, \citealt{2017AJ....154..108J,2017AJ....154..107P}), in conjunction with improved parallaxes from Gaia \citep{2017AJ....154..107P,2018MNRAS.479.4786V}, has revealed a distinct ``radius valley'' \citep{2017AJ....154..109F} among the population of planets hosted by stars. 
This radius valley of $1.5-2R_{\oplus}$ \citep{2017MNRAS.Owen,2020MNRAS.493..Gupta,2022MNRAS.Gupta} separates smaller super-Earths from larger sub-Neptunes. 
The investigation of super-Earth formation could facilitate the identification of the physical mechanism of the bimodal structure of planet radii.

The formation of super-Earth could be classified into two categories based on the material distribution of protoplanetary disks. 
The suppression of runaway accretion within the framework of core accretion can lead to the formation of either a super-Earth or a mini-Neptune when the planet is still embedded within a gas-rich protoplanetary disk \citep{2013MNRAS.Chiang,2015ApJ...Lee,2017ApJ...850..198Y,2020MNRAS.Ali,2021ApJ...Zhong,2023arXiv230605913J}. 
In a disk with low gas content, planets generally experience an atmospheric loss due to either core-powered atmospheric escape \citep{2018MNRAS.Ginzburg,2020MNRAS.493..Gupta,2022MNRAS.Gupta} or photo-evaporation \citep{2014ApJ...Kurokawa,2017MNRAS.Owen,2023MNRAS...Owen}.

The giant impact potentially serves as another viable mechanism for planetary atmospheric escape  (\citealt[hereafter BS19]{2019..Biersteker..impacts}; \citealt{2019Natur.Liu}). The mechanical and thermal effects resulting from a giant impact could trigger the formation of the Parker wind, leading to the loss of atmospheric mass and the expansion of the envelope (BS19). Current studies primarily focus on the hydrodynamic consequences of a giant impact \citep{2019Natur.Liu}, and further studies are necessary to understand the effects of atmospheric loss in post-impact planets on the long-term planet formation process. The atmospheric loss during the post-impact stage is influenced by the separation between the planet and its host star and its initial thermal state. 
Younger planets closer to their host stars are more susceptible to impact-driven atmospheric loss (BS19). When the planet envelope's self-gravity is considered, the planet requires more energy input or additional impacting bodies to achieve complete mass loss \citep{2022RAA....Huang}.

The dynamical interplay between planets and surrounding protoplanetary disks leads to orbital migration   (Type-I migration, \citealt{1980ApJ...241..425G,1997Icar..126..261W, 2008ApJ...673..487I}) and eccentricity attenuation. Nevertheless, the dispersion of the disk would nullify its damping effect, leading to instability of the planet's orbit due to the increase of planet eccentricities by orbit crossing and giant impact events \citep{2014A&ACossou,2019Denham}. 
Typically, short-period super-Earths are situated either at the periphery of the Roche lobe  of the planet or in the process of migrating towards that particular region \citep{2016CeMDA.Jackson}. The planet's orbital decay in a star-planet system is attributed to the tidal torques. This is particularly evident when a planet is in close proximity to its host star \citep{2016CeMDA.Jackson,2017MNRAS...Ginzburg}. Furthermore, the planet's atmospheric escape may alter the orbital angular momentum and expand the planet's orbital semi-major axis. 

The prevailing approach for investigating orbital migration induced by mass loss primarily involves photo-evaporation \citep{2012Jackson,2016CeMDA.Jackson}. The increase of the orbital semi-major axis is directly related to the rise of the mass loss rate. \cite{2012MNRAS.425.Owen,2017ApJ...847...Owen,2022ApJ...928..105F...Fujita} proposed that the photo-evaporation effect is more pronounced for planets that are closer to the central star. The displacement of a planet will be significant when its proximity to the central star leads to a substantial increase in its mass-loss rate. In particular, photo-evaporation processes would form a radius valley for close-in planets \citep{2022ApJ...928..105F...Fujita}.
\cite{2014ApJ...Valsecchi} examined the conditions under which gas giants that experience orbital migration due to photo-evaporation could survive.
However, there is a lack of studies on the migration of post-impact planets caused by impact-driven mass loss. 

The equilibrium temperatures of planets, categorized as hot, warm, or cold, are subject to modification due to ongoing mass loss, consequently affecting the distribution of planetary radii \citep{2018MNRAS.Ginzburg}. The equilibrium temperatures are predominantly influenced by the host star, implying that the radius distribution of planets is also affected by various spectral types of host stars. The Kepler mission has confirmed that many super-Earths and sub-Neptunes orbit low-mass stars with the FGKM spectral type \citep{2018AJ...Fulton,2018MNRAS.Van...Eylen,2021MNRAS.Van...Eylen}. Studies have suggested that the star's spectral type played a role in the orbital migration resulting from photo-evaporation \citep{2022ApJ...928..105F...Fujita}. 

Compared to the mechanical effect, the heating effect of a giant impact is more important (BS19). Therefore, our main focus is on the thermal consequences resulting from such events. In terms of thermal effects, planetesimals are more likely to cause impacts during this period as their mass loss is significantly less than that of the target planets. It should be noted that giant impacts are relatively rare occurrences. In this work, whenever we mention being ``impact/collision", we specifically refer to planetesimal impacts in future scenarios.

Prior works primarily concentrate on the radius valley driven by the photo-evaporation. We instead dwell on the planet's radius distribution subject to mass loss and orbital migration induced by impacts. 
After the collision, planets orbiting F, G, and K-type stars experience different outcomes due to the interplay between mass loss and orbital migration. As the host star transitions from a K-type to an F-type, its temperature, radius, and luminosity increase. This leads to an increase in the planet's outer temperature and a decrease in radius, which affects mass loss and orbital migration. In addition, the efficiency of  the planetesimal  impact is influenced by both the planet's initial gas-to-core mass ratio (GCR)
and the impactor's mass.  Smaller (low-GCR) planets experience a significantly greater loss of atmosphere upon impact by the same object compared to massive planets. However,  massive planets can retain more atmosphere due to the increase in gravitational energy and the suppression of mass loss caused by orbital migration. By combining these signatures, it is possible that planetesimal impacts could facilitate the formation of a radius valley.

This paper is structured as follows: \S  \ref{sec2} describes the planet's mass loss and the orbital evolution following an impact event. \S \ref{sec_result}  presents that planets orbiting host stars of various spectral types undergo the planetesimal impact mechanism, resulting in the formation of a radius valley. 
The subsections in this section are as follows: \S \ref{sec:result1} 
 explores core energy-dominated mass loss, factors determining complete mass loss due to collisions, and the influence of the initial orbit's semi-major axis on mass loss. \S \ref{sec:host_spectral_type} investigates planet mass loss around different host stars and its role in forming radius valleys.
Finally, in \S \ref{section_conclusion}, we summarize and discuss our calculation results.

\section{The mass loss and migration of the post-impact planet}
\label{sec2}
Note that the planet's mass loss may influence its orbital evolution. The orbital evolution is intricately related to its structural, energy adjustment and mass loss.
We describe the planet structure undergoing mass losses in \S \ref{sec2_1:structure}. In \S \ref{sec2_2:orbital_evolution}, we outline the planet's orbital evolution due to the mass loss.

\subsection{The Thermal Structure of Post-impact Planets}
\label{sec2_1:structure}

The internal structure of super-Earths is governed by the equations of mass conservation, hydro-static equilibrium, energy transport, and energy conservation \citep{2013sse..book.....Kippenhahn}
\begin{equation}
    \frac{\mathrm{d} M_{r}}{\mathrm{d} r} = 4 \pi r^2 \rho \ ,
    \label{eq1:mass_conservation}
\end{equation}
\begin{equation}
    \frac{\mathrm{d} P}{\mathrm{d} r} = -\frac{GM_{r}}{r^2}\rho \ ,
    \label{eq2:hydrostatic_equ}
\end{equation}
\begin{equation}
    \frac{{\rm d} T}{{\rm d}r} = \frac{T}{P}\frac{{\rm d}P}{{\rm d} r} \nabla \ ,
    \label{eq3:energy_transfer}
\end{equation}
\begin{equation}
    \frac{{\rm d}L}{{\rm d}r} = \frac{{\rm d}M_{r}}{{\rm d}r} \epsilon_{\rm g} \ ,
    \label{eq4:energy_conservation}
\end{equation}
where the variables $M_{r}$, $P$, $T$, $\rho$, $L$, and $r$ denote the planet enclosed mass within radius $r$, atmospheric pressure, temperature, density, luminosity, and radial distance, correspondingly. 
And $G$ is the gravitational constant. 
Meanwhile, the energy released by gravitational contraction is $\epsilon_{g}$.
The temperature gradient $\nabla$ is determined by the minimum of the adiabatic gradient $\nabla_{\rm ad} = (\gamma-1) / \gamma = 2/7$ ($\gamma$ is the adiabatic index of the envelope) and the radiative gradient $\nabla_{\rm rad}$, i.e., $\nabla=\min\left(\nabla_{\rm ad}, \nabla_{\rm rad}\right)$. 
The radiative gradient satisfies 
\begin{equation}
    \nabla_{\rm rad} = \frac{3\kappa P L}{64 \pi \sigma G M_{r} T^4} \ ,
    \label{eq5:radiative_gradient}
\end{equation}
where the parameters $\kappa$ and $\sigma$ correspond to the opacity and the Stefan-Boltzmann constant, respectively.
Follow BS19, we set $\kappa = 0.1 cm^{2}/g$.

The outer boundary radius $R_{\rm out}$ of a planet depends on the smaller value between the Bondi radius $R_{\rm B}$ and the Hill radius $R_{\rm H}$, i.e., $R_{\rm out} = \min \left(R_{\rm B}, R_{\rm H} \right)$ \citep{2014..Lee..ApJ...797...95L}. 
In general, the Bondi radius and Hill radius correspond to $R_{\rm B} = G M_{\rm p}/c_{\rm s}^2$  and $R_{\rm H} = a\left[M_{\rm p}/3\left(M_{\rm p}+M_{\star}\right)\right]^{1/3}$, where the symbols $M_{p}$ and $M_{\star}$ are the total mass of the planet and the mass of the central star, respectively. 
The sound speed $c_{\rm s} = \sqrt{P_{\rm out}/\rho_{\rm out}} =\sqrt{\gamma k_{\rm B}T_{\rm out}/\mu} $ depends on the outer pressure $ P_{\rm out}$ and density $\rho_{\rm out}$. 
The temperature at the outer boundary $T_{\rm out}$ is always equal to the equilibrium temperature $T_{\rm e q}$, which satisfies 
\begin{equation}
T_{\rm e q}=\left(\frac{1-A_B}{4}\right)^{\frac{1}{4}}\left(\frac{R_{\star}}{a}\right)^{\frac{1}{2}} T_{\star} \ ,
\label{eq:T_eq}
\end{equation}
where $A_B = 0.29$, and $a$, $R_{\star}$ and $T_{\star}$ are the Bond albedo, planet orbital semi-major axis, the stellar radius and temperature of the central stellar, respectively. 

The planet's mass loss depends on its energy, including its gravitational and thermal energies (BS19), i.e., $E= E_{\rm grav}+E_{\rm th}$. 
Generally, the parameters at the radiative-convective boundary (RCB) can be explored to obtain the planet's gravitational energy accurately.
The gravitational energy is shown as follows 
\begin{equation}
E_{\rm grav}=-4 \pi G \int_{R_{\mathrm{c}}}^{R_{\mathrm{rcb}}} M_{\mathrm{r}}  \rho r \mathrm{~d} r 
 \ ,
\end{equation}
where the subscript ``${\rm rcb}$" represents the parameters on the RCBs.
Meanwhile, thermal energy is 
\begin{equation}
E_{\mathrm{th}}=\frac{4 \pi}{\gamma-1} \int_{R_{\mathrm{c}}}^{R_{\mathrm{rcb}}} P r^2 \mathrm{~d} r \,.
\end{equation}

The Super-Earths are postulated to possess a fully molten core, wrapped by primordial hydrogen-helium envelopes, devoid of a solid insulating outer layer \citep{2022ApJ...928..105F...Fujita}. 
The temperature at the base of the envelope closely approximates the core temperature, $T_{\rm b} \approx T_{\rm c}$, due to efficient heat transfer between the core and the deep region of the envelope.
Additionally, it is assumed that the initial core temperature surpasses 2000 K and that the core energy adheres to the following equation 
\begin{equation}
E_{\mathrm{c}} \sim c_{ \mathrm{v,c}} M_{\mathrm{c}} T_{\mathrm{c}}\,,
\label{eq8:CoreEnergy}
\end{equation}
where the specific heat capacity of the core
$c_{\rm v,c} \sim 5-10 \times 10^6 {\rm erg} \,\, {\rm g}^{-1} K^{-1}$
\citep{1995ApJ...450..463G..Guillot}. 
In summary, a planet's total energy encompasses its core, gravitational, and thermal energy, that is, 
\begin{equation}
    E_{\rm total} = E_{\rm c} + E_{\rm grav} + E_{\rm th} \,.
    \label{eq9:TotalEnergy}
\end{equation}

Typically, the atmospheric mass loss limited by Bondi radius can be delineated by the following equation \citep{2018MNRAS.Ginzburg,2022ApJ...928..105F...Fujita,2022MNRAS.Gupta}
\begin{equation}
\dot{M}_{\rm env}^{\rm B}=-4 \pi \rho_{\mathrm{s}} c_{\mathrm{s}} r_{\mathrm{s}}^2=-4 \pi r_{\rm s}^2 c_{\rm s} \rho_{\rm r c b} \exp \left(-\frac{G M_{\rm p}}{c_{\rm s}^2 R_{\rm r c b}}\right)\,,
\label{eq10:mass_loss}
\end{equation}
where the subscript ``${\rm s}$" represent the parameters at the sonic point. 
Meanwhile, the parameters at the sonic point change with the temperature, density, and radius specified by the outer boundary.

The atmosphere diffuses outward from the RCB, while the planet maintains a stable radiative zone through its cooling and the energy supplied by its host star. 
If the planet's cooling luminosity prevails, the upper limit of the mass-loss rate of its hydrogen-helium envelope (BS19) is denoted by 
\begin{equation}
\dot{M}_{{\rm e n v,} \max } \approx-\frac{L_{\rm rcb} R_{\rm rcb}}{G M_{\rm c}}\,.
\label{eq12:max_mass_loss_rate}
\end{equation}
The final mass-loss rate of a post-impact planet is the minimum value of $\dot{M}_{\rm env}^{\rm B}$ and $\dot{M}_{\rm env, max}$\,.

Once the adiabatic index $\gamma$ exceeds $4/3$, most atmospheric mass is concentrated within the interior convective region. 
As a result, elevating a gaseous parcel from the RCBs to an infinite height needs a corresponding alteration in the envelope's energy \citep{2022MNRAS.Gupta}
\begin{equation}
\dot{E}_{\mathrm{env}, \mathrm{m}} \approx \frac{G M_{\mathrm{c}} \dot{M}_{\mathrm{env}}}{R_{\mathrm{r c b}}}\,.
\label{eq11:massloss_energy_m}
\end{equation}
The luminosity at the RCB can be estimated through the integration of the flux conservation and hydrodynamic equilibrium equations. Consequently, the energy alteration represented by luminosity is expressed as (BS19)
\begin{equation}
\dot{E}_{\mathrm{env}, \mathrm{L}}=-L_{\mathrm{rcb}}=-\nabla_{\mathrm{ad}} \frac{64 \pi \sigma T_{\mathrm{rcb}}^3 G M_{\mathrm{c}} \mu}{3 \kappa_{\mathrm{R}} \rho_{\mathrm{rcb}} k_{\mathrm{B}}}.
\label{eq13:energy_l}
\end{equation}
By combining Equations (\ref{eq9:TotalEnergy},\ref{eq11:massloss_energy_m}-\ref{eq13:energy_l}), the timescale between two consecutive snapshots can be derived as 
\begin{equation}
    \Delta t = \frac{E_{\rm c} + E_{\rm grav} + E_{\rm th}}{\dot{E}_{\mathrm{env}, \mathrm{L}} + \dot{E}_{\mathrm{env}, \mathrm{m}}}\,.
    \label{eq12:}
\end{equation}

Collisions result in various consequences, such as hit-and-run incidents, merging events, partial accretion and erosion processes, as well as catastrophic disruptions \citep{2012ApJ...Leinhardt,2012ApJ...Stewart}. 
As indicated in Table 1 of \cite{2012ApJ...Stewart}, giant impacts in collision tests are less frequent compared to planetesimal impacts which occur more frequently. Considering the thermal effects of giant impacts as our main concern, the size of planetesimals better aligns with the requirement for a complete impact when considering thermal effects rather than mechanical effects.  When the impact velocity is approximately 1-2 times the escape velocity and the impactor mass is equal to 1/40 times the target planet mass 
\citep{2012ApJ...Stewart}, only 4\% of all planetesimal impact during the formation of Earth-mass planets are accounted for by graze-and-merge, while hit-and-run makes up 24\%.  In addition, approximately 70\% of planetesimal impacts primarily involve partial accretions.  Although these probabilities increase as the planet's mass decreases, our main focus is on investigating the process of partial accretion following a planetesimal impact and its subsequent energy input into a target planet with a core mass of $5M_{\oplus}$.  For our planetesimal impact model, we assume an impact velocity ($v_{\rm imp}$) similar to escape velocity ($v_{\rm eac}$).


This work primarily focuses on the thermal effects of the planet, rather than delving into the mechanical shock effects caused by the collision. 
BS19 suggested that the thermal effects of planetesimal impact are considerably more significant than those caused by mechanical shock.
The impactor's kinetic energy can be converted into thermal energy and deposited within the envelope and core of the planet. 
Note that the planetesimal impact can sufficiently heat both the core and envelope when the impact velocity is comparable to the escape speed of the planet, i.e., $v_{\rm imp}\sim v_{\rm esc}$. The impact energy should satisfy 
\begin{equation}
E_{\rm imp} =  \eta M_{\rm imp} \frac{v_{\rm imp}^2}{2} =  \eta M_{\rm imp} \frac{G M_{\rm p}}{R_{\rm c}} \ ,
\label{eq:impact_energy}
\end{equation}
where $M_{\mathrm{imp}}$ denotes the impactor's mass. The parameter $\eta \in[0,1]$ is the fraction of the total impact energy which is available for heating the core and envelope.
When the impact is considered, the impact energy should be added to the numerator of Equation (\ref{eq12:}). 

\subsection{The Orbital Evolution}
\label{sec2_2:orbital_evolution}

The semi-major axis of the mass-loss planet evolves due to tidal effects, which cause it to move outward in order to compensate for the variation in the angular momentum. The sonic point of the planet is determined by
\citep{2022ApJ...928..105F...Fujita}
\begin{equation}
2 c_{\mathrm{s}}^2=\frac{G M_{\mathrm{p}}}{r_{\mathrm{s}}}-\frac{3 G M_{\star} r_{\mathrm{s}}^2}{a^3} \ ,
\label{eq:radius_at_sound_speed}
\end{equation}
where the central star's mass is $M_{\star}$. For simplicity, the post-impact planet is assumed to lose its mass in an isotropic manner. 
The orbital angular momentum of the star-planet system is
\begin{equation}
J
=M_{\star} M_{\mathrm{p}} \sqrt{\frac{G a}{M}} \ ,
\end{equation}
where the total mass of the star–planet system is $M=M_{\star} + M_{\rm p}$.
The time derivative of orbital angular momentum reads 
\begin{equation}
\frac{\dot{J}}{J}=\frac{\dot{M}_{\star}}{M_{\star}}+\frac{\dot{M}_{\mathrm{p}}}{M_{\mathrm{p}}}+\frac{1}{2} \frac{\dot{a}}{a}-\frac{1}{2} \frac{\dot{M}}{M} \ .
\label{eq:dot_J}
\end{equation}

The planet orbital migration is determined by the variations of the orbital angular momentum of the star-planet system \citep{2022ApJ...928..105F...Fujita}. 
It is hypothesized that the Parker wind, generated subsequent to a significant impact, will disperse the planetary matter during the process of planetary migration.
 If $\dot{J} = 0$ and $\dot{M}_{\star} = - \dot{M}_{\rm p}$ in Equation (\ref{eq:dot_J}), the conservative mass transfer from a planet to its host star correspond to $\dot{a}/a = - 2\dot{M}_{\rm p} \left(M_{\star} - M_{\rm p}\right)/\left(M_{\star} M_{\rm p}\right)$ \citep{2022ApJ...928..105F...Fujita}. In addition, the change in angular moment satisfies $\dot{J} = \dot{M}_{\rm p} a_{\rm p}^2\Omega$ for the atmosphere escape.

Random occurrences of the planetesimal impact events, whether direct or oblique, result in the formation of Parker winds. These winds are generated by a planet and give rise to a comet tail or a rapid ring of hydrogen atoms through the exchange of charges between escaping hydrogen atoms and protons in the stellar wind.
In general, the change in the angular momentum of a star-planet system is assumed to be  $\dot{J}=(1-\chi) \dot{M}_{\mathrm{p}} a_{\mathrm{p}}^2 \Omega$, where $0 \leqslant \chi \leqslant 1 $ is the fraction of angular momentum conserved in the system and the parameter $\Omega$ is Keplerian angular velocity \citep{2022ApJ...928..105F...Fujita}.  
Note that the parameter $\chi$ may vary with time due to the geometry of the outflowing wind, $\chi$ is assumed to be constant in this study.
The detailed behavior and geometry of the escaping atmosphere are also not considered in this study. 
According to Equation (\ref{eq:dot_J}), the orbital migration satisfies 
\begin{equation}
\frac{\dot{a}}{a}=-2 \chi \dot{M}_{\mathrm{p}} \frac{M_{\star}-M_{\mathrm{p}}}{M_{\star} M_{\mathrm{p}}}-(1-\chi) \frac{\dot{M}_{\mathrm{p}}}{M}>0 \ .
\label{eq:migration_conditions}
\end{equation}
Given the planetary core mass, the mass loss of the planet  $\dot{M}_{\mathrm{p}}$ approximates the mass loss of its atmospheric envelope $\dot{M}_{\rm env}$.

\section{Numerical Results}\label{sec_result}

We present our numerical results of the mass loss and orbital migration of post-impact planets.  In \S \ref{sec:result1}, we show how different envelope base temperatures and initial semi-major radii affect both planetary mass loss and orbital evolution. In addition, the decisive factors of complete mass losses are also included.
In \S \ref{sec:host_spectral_type}, we concentrate on the influences of different spectral types of host stars on the planetary radius distribution.

\subsection{The Mass Loss}
\label{sec:result1}

We follow the method mentioned in Appendix \ref{method} for the planet's mass and orbital evolution. In \S \ref{section:a_change_t}, we explore 
the mass loss for 
planets with different base temperatures (i.e., $T_{\rm b,0}$). Subsequently, we show the outcomes of the orbital migration and 
the fraction of the planet's H/He envelope lost (i.e., $X$) at the different scenarios in \S \ref{3.1.3_the_impact_mass}. Finally, we test the effects of the initial semi-major radii on  the mass loss 
of the post-impact planet in \S \ref{sec3.2}. 

To be specific, we set gas-to-core ratio $GCR=0.03$, the initial mass as $M_{\rm env,0} = 0.03 \,M_{\rm c}$ \citep{2022ApJ...Izidoro}, the initial semi-major axis $a_{\rm initial } = 0.1 AU$, the core mass is $M_{\rm c} = 3M_{\oplus}$ with the core radius $R_{\rm c} = \left(M_{\rm c}/M_{\oplus}\right) = 1.32 R_{\oplus}$ in this section \citep{2006Icar..181..545V..Valencia}. 
The opacity is denoted as $\kappa = 0.1 \,\rm g^{-1}\, cm^{2}$. 
The typical physical parameters of the Sun are adopted for the central host star and the parameter $\chi$ is assigned a value of 0.5 \citep{2022ApJ...928..105F...Fujita}.

\subsubsection{Effects of Initial Thermal States}
\label{section:a_change_t}

\begin{figure}   
         \includegraphics[width=9cm]{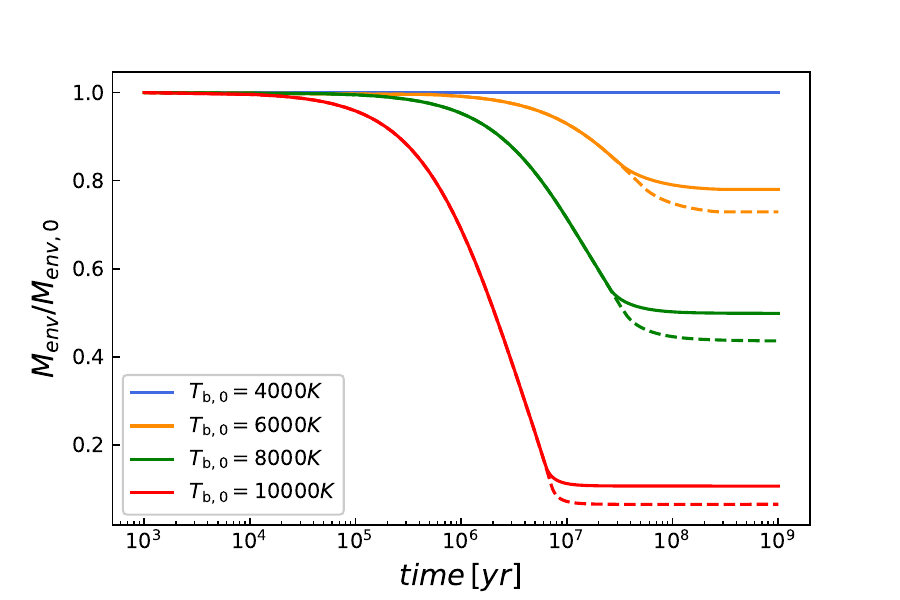}
    \caption{
             The mass evolution for the planets with various base temperatures.
            The red, green, orange, and yellow colors represent the base temperatures of $10,000K$, $8,000K$, $6,000K$, and $4,000K$, respectively. 
            Solid lines denote the case of the planet with the orbital evolution,  while dashed lines show the non-migration case.}
    \label{fig1:without_giant_impact}
\end{figure}

Note that the mass loss of a close-in evaporating planet may alter its angular momentum, resulting in outward migration and an increase of semi-major axis to maintain angular momentum equilibrium \citep{2022ApJ...928..105F...Fujita}. Actually, the mass loss by  collision could facilitate orbital migration as well. 
Therefore, in this work, we utilize orbital migration as an indicator of planetary mass loss.
The mass loss relies upon the planet's orbital semi-major axis and initial thermal state (BS19). 
The initial thermal state dictates the planet's base temperature (i.e., $T_{\rm b,0}$), thereby influencing the magnitude of mass loss. Figure \ref{fig1:without_giant_impact} presents the planet's mass loss.  
The planet mass loss without orbital migration is shown by dashed lines in the upper panel. The behavior of mass loss is a bit similar to the evolution\footnote{We include planet self-gravity in our numerical calculations. For a more massive planet gas envelope, the difference between our results and those in BS19 would be more evident.} illustrated in Figure 3 of BS19. 
As mentioned in \S \ref{sec2_1:structure} on planetary structure, the total energy of a planet comprises core-powered energy, heat energy, and gravitational potential energy. Sufficient energy is required for a planet to overcome gravity and facilitate atmospheric escape. 
In general, when the sum of the core-power energy and thermal energy of the H/He envelope exceeds the gravitational potential energy, $E_{\rm c}+E_{\rm t h}>E_{\rm grav}$, the mass loss will be triggered. 
The core-powered energy increases with $T_{\rm b,0}$, providing the planet with sufficient energy to overcome gravity and lose its mass completely.

Different from BS19, we have considered the planet's orbital migration induced by mass loss. The outward migration of the planet effectively offsets alterations in its angular momentum \citep{2022ApJ...928..105F...Fujita}. The initial configuration of the migrating planet remains consistent with the dashed lines, while the mass-loss rate is primarily influenced by the planet's structure rather than core-power energy.  
Especially, the outer boundaries of the planet's evolution are subject to modification due to orbital migration.

The planet migrates away from its host star when it loses its atmosphere. 
During the expansion phase of the planet, there is a  
augmentation in $R_{\rm out}$ as it migrates away from its original orbit. The temperature at the outer boundary diminishes, resulting in a decrease in $T_{\rm rcb}$ due to the relationship $T_{\rm rcb} \sim T_{\rm eq} \propto a^{-1/2}$.
Furthermore, the sound speed decreases as $T_{\rm eq}$ is reduced, which can be expressed as $c_{\rm s} \propto T_{\rm eq}^{1/2}$. To accommodate the changes in $c_{\rm s}$ and $a$, the radius at the sonic point ($r_{\rm s}$) in Equation (\ref{eq:radius_at_sound_speed}) will slightly increase.   However, this increase can be disregarded.

We integrate Equation (\ref{eq3:energy_transfer}) from the core outward to obtain the radial density distribution of the convective zone, shown as follows
\begin{equation}
    \frac{T}{T_{\mathrm{b}}}=\nabla_{\rm ad} \frac{G M_{\mathrm{c}} \mu }{R_{\mathrm{c}} k_{\mathrm{B}} T_{\mathrm{b}}}\left(\frac{R_{\mathrm{c}}}{r}-1\right)+1 \ ,
\label{eq:t_profile}
\end{equation}
since the planet's total mass is primarily composed of the core mass, that is, $M_{\rm p} \approx M_{\rm c}$. Meanwhile, $R_{\rm rcb}$ increase with the reduction in $T_{\rm rcb}$ in Equation (\ref{eq:t_profile}) under the same conditions.
The pressure $P\left(\rho, T\right)$ in the adiabatic zone satisfies the following relationship $P = K \rho^{\gamma} = \rho k_{\rm B} T /\mu$ with the polytropic constant $K = P_{\rm rcb}/\rho_{\rm rcb}^{\gamma}$.
Combined with Equation (\ref{eq:t_profile}), we can obtain the radial density distribution in the convective zone as follows
\begin{equation}
\frac{\rho}{\rho_{\mathrm{b}}}=\left[\nabla_{\rm ad} \frac{G M_{\mathrm{c}} \mu }{R_{\mathrm{c}} k_{\mathrm{B}} T_{\mathrm{b}}}\left(\frac{R_{\mathrm{c}}}{r}-1\right)+1\right]^{\frac{1}{\gamma-1}} \ .
\label{eq:density_in_adibatic}
\end{equation}
Thus, $\rho_{\rm rcb}$ is decreased due to the increase of $R_{\rm rcb}$.

The actual rate of atmospheric mass loss is subject to two limitations, namely $\dot{M}_{\rm env}^{\rm B}$ and $\dot{M}_{\rm env, max}$, both of which are influenced by parameters on the RCB. 
The change in $\dot{M}_{\rm env}^{\rm B}$ depends on several parameters, such as $c_{\rm s}$, $\rho_{\rm rcb}$ and $R_{\rm rcb}$. Since the decrease of $c_{\rm s}$ can counteract the effect of the increase of $R_{\rm rcb}$, the mass loss rate will decrease, along with other parameters. Meanwhile, 
$\dot{M}_{\rm env, max} \propto R_{\rm rcb}L_{\rm rcb} \propto T_{\rm rcb}^{3} R_{\rm rcb}/\rho_{\rm rcb}$. The envelope of an atmospheric escaping planet is much thinner, with minimal variability in $\rho_{\rm rcb}$. In addition, the decrease of $T_{\rm rcb}^{3}$ tends to be more significant than the decrease of $R_{\rm rcb}$. Finally, the value of $\dot{M}_{\rm env, max}$ also declines. In summary, the actual mass-loss rate is reduced by the orbital migration.

The rate of orbital migration is proportional to the mass loss \citep{2022ApJ...928..105F...Fujita}. The expansion of the orbital semi-major axis will diminish the planet's mass loss rate. These two coupled processes ultimately stabilize the planet's orbital semi-major axis at a constant value. Consequently, the mass of the planetary retention envelope (solid lines) is relatively greater than the non-orbital migration (dashed lines) cases.
With a base temperature of 4000K, the planet would not lose mass and migrate since its energy is insufficient. 
Planets with higher base temperatures experience more substantial mass loss and migrate a greater distance. 
Note that the initial thermal state of a planet is related to its age, younger planets with a higher $T_{\rm b,0}$ demonstrate a heightened propensity for substantial mass loss and orbital migration.

\subsubsection{The Decisive Factors of Complete Mass Losses}
\label{3.1.3_the_impact_mass}

\begin{figure}
        \includegraphics[width=9cm]{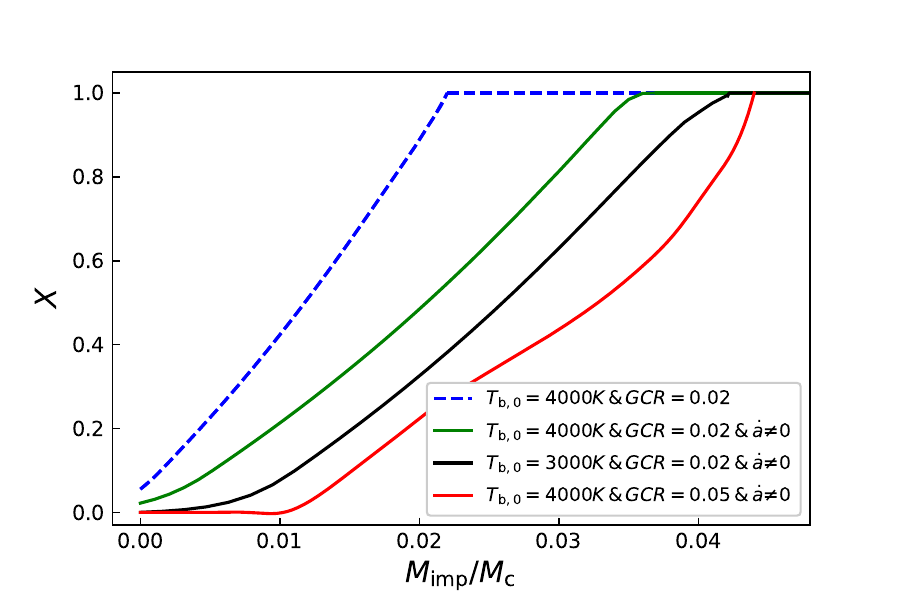}
         \includegraphics[width=9cm]{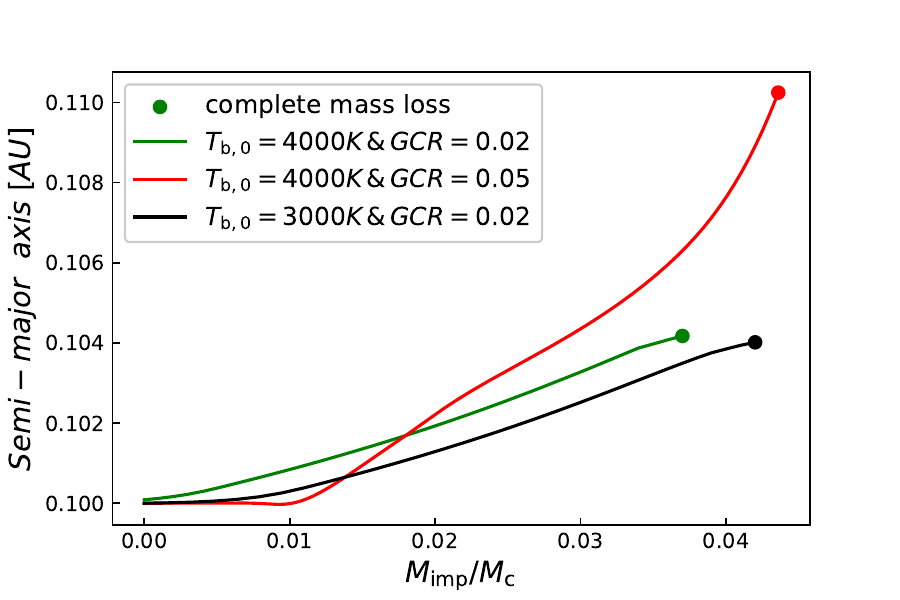}
    \caption{
    The upper panel: the variation of the fraction of the H/He envelope lost ($X$) of a planet with the different impactors' masses. 
    The lower panel:  The orbital migration of the super-Earth with the different impactor's masses.   The final semi-major axis after migration is finished.
    The blue dashed line displays the case without migration, with an initial base temperature of $T_{\rm b,0}=4000$K and a gas-to-core ratio of $\rm GCR=0.02$.
    In contrast, the solid green line represents the scenario with the planetary orbital migration. 
    The solid black line further displays the planetary migration scenario with $T_{\rm b,0}=3000$K.
    The solid red line, on the other hand, presents a scenario where the super-Earth migrates, with  $T_{\rm b,0}=4000$K and $\rm GCR=0.05$. 
    }
    \label{fig2:example_figure}
\end{figure}

This section studies how the impactor's mass affects the planetary  
mass-loss process. The mass loss subsequent to a giant impact is predominantly attributed to core-powered heating. 
The upper panel of Figure \ref{fig2:example_figure} illustrates the fraction of the H/He envelope lost after 2 Gyr due to collisions with impactors of varying masses. Meanwhile, the corresponding orbital migration is shown in the lower panel. The dots denote the planet reaching a complete mass loss. 

The in-situ mass loss of the planet, shown by the blue dashed line in Figure \ref{fig2:example_figure}, is considered as a reference for comparison. The mass evolution without orbital migration is consistent with Figure 5 in BS19.  As elucidated in \S \ref{section:a_change_t}, the mass loss depends on the energy from the core. As this energy increases with the mass of the impact object ($M_{\rm imp}$), the value of $X$ also increases. When $X=1$, a complete mass loss develops. Nevertheless, when accounting for orbital migration, the trajectory undergoes alteration.  As depicted by the lower panel's green line, the change in angular momentum intensifies with the mass of the impacting body ($M_{\rm imp}$), leading to greater atmosphere retention by the planet compared to the in-situ scenario. Consequently, the final value of $X$ decreases as indicated by the green line. Simultaneously, the critical mass of the impacting body required for complete mass loss increases with the outer radius, i.e., $M_{\rm imp, crit} \propto 1-R_{\rm rcb,0}/R_{\rm out}$,  while maintaining the same initial radius at the RCB ($R_{\rm rcb,0}$) in an evolution.  
Therefore, a migrating planet necessitates a significant great mass of the impact object to enhance the core-powered energy and then lose its mass completely. 


The mass loss and orbital migration are 
amplified in younger planets possessing a less substantial envelope. As mentioned in  \S \ref{section:a_change_t}, the decrease in a planet's core energy prior to collision leads to a decrease in the available energy to counteract gravity, resulting in a decrease in mass loss and orbital movement. As a result, the planetary atmospheric envelope possesses a greater final mass than the reference model, leading to a reduction in the planet's $X$ parameter. Furthermore, the final semi-major axis experiences a decrease. However, augmenting the mass of the impacting object may indirectly amplify the core energy. Hence, a planet with diminished core energy necessitates a more massive impactor to shed its envelope entirely. Furthermore, the energy retained by the planet after overcoming gravitational potential energy is increased, which will augment the $X$ and $a$ parameters.

A planet's orbit can be significantly affected by the strength of its initial GCR depicted by the red lines. 
Seen in Appendix B, the effect of self-gravity increases with the initial GCR (i.e., initial envelope mass).
As a planet's atmospheric mass increases, its gravitational force becomes stronger, thus making it more challenging for the planet to lose mass and migrate outwards. 
Consequently, the final $X$ and orbital semi-major axis experience a decline. 
To achieve complete mass loss, planets require larger critical impactor masses to overcome the challenges own orbital migration and increased gravitational energy. 
The final outcome shows that these planets can lose all their mass and be propelled to the furthest point, causing a significant impact on planetary orbital migration.
Therefore, collisions between massive planets and impacting objects have a significant effect on planetary orbital migration.
In addition, suitable {\bf planetesimal} impacts could transform Sub-Neptunes into super-Earths with bare cores and make them migrate further.

The augmentation of core energy within a planet exhibits a proportional relationship with the mass of the impactor. In cases where the initial core energy of the planet is lesser or the initial  GCR is substantial, the internal core energy can be elevated by augmenting the mass of the impactor. Consequently, it becomes imperative to select a suitable impact mass when investigating the mass loss and orbital migration of planets orbiting host stars of diverse spectral types.

The orbital migration distance is relatively small. We think that planetary migration is a direct result of mass loss and serves as an indicator of its commencement. The likelihood of a planet undergoing migration is directly proportional to its ability to shed mass. In this study, we examine the effect of impactors on the mass loss of planets orbiting different central stars. When using the same impactor mass to determine the mass loss following the impact of planets with varying masses after 2 Gyr, some planets exhibit greater gravitational attraction, requiring a higher level of impact energy to induce mass loss. This suggests that these planets are resistant to mass loss. Hence, the orbital migration is not readily apparent.  However, 
when studying the relationship between radius and orbital period after evolution, although the role of mass loss-induced orbital migration is small, it remains a valuable indication of the orbital period for planetary mass loss in this work. The distribution of planet radius versus orbital period would be more realistic when planet migration is considered.


\subsubsection{Effect of the Initial Semi-major Radius}
\label{sec3.2}

\begin{figure}
    \includegraphics[width=9cm]{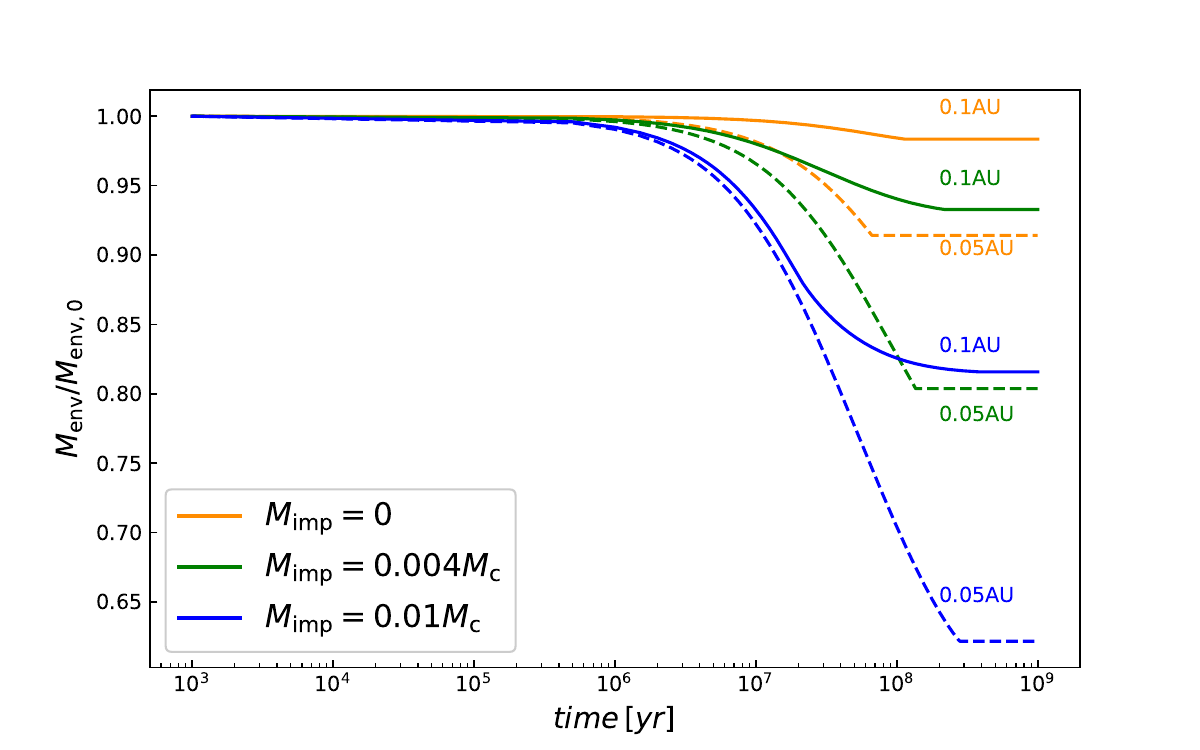}
    \caption{
     The effect of the planetesimal impact on a planet's mass evolution.
    The orange line is the case without 
     collision.
    Note that the green and blue lines indicate the different impacting masses. 
     The solid and dashed lines correspond to cases of the planet at 0.1 AU and 0.05 AU, respectively. 
    }
    \label{fig3:evapration_impact}
\end{figure}

The fiducial model without the planetesimal impact adopts $T_{\rm b,0} = 4000\ K$ and $\rm GCR = 0.025$, represented by the solid orange line in Figure \ref{fig3:evapration_impact}. The mass loss in this model is solely determined by its initial core-powered energy. Note that this energy is insufficient to cause significant mass loss. 
However, the planetesimal impact tends to increase 
mass loss 
compared to the fiducial model. Furthermore, the mass of the impacting object is a critical factor in determining 
 the mass loss. 
In Figure \ref{fig3:evapration_impact}, the green and blue solid lines correspond to impacts with impactor masses of $0.004M_{\rm c}$ and $0.01M_{\rm c}$, respectively. Smaller impactors lead to a slight increase in 
mass loss, 
while larger impactors result in a significant rise in the planet's core energy. This leads to considerable increases in 
mass loss. 
Therefore, an appropriately sized impactor can achieve 
complete mass loss within 0.1 AU.

The initial semi-major axis of a planet's orbit is closely linked to 
he impact-driven mass loss. 
\S \ref{section:a_change_t} explains that when the semi-major axis decreases, it can lead to changes in the planet's interior structure and intensify these phenomena. This effect is more pronounced compared to the fiducial model, resulting in post-impact planets having a lighter final mass.
Our numerical results show that the appropriate impactor mass would significantly increase the mass loss. 
Moreover, the effect of the planetesimal impact is also significant at a distance of 0.1 AU. As such, we will examine whether the planetesimal impact can contribute to the formation of the radius valley at this distance.

\subsection{ Effects of Host Star Spectral Types}
\label{sec:host_spectral_type}

\begin{figure}
    \includegraphics[width=9cm]{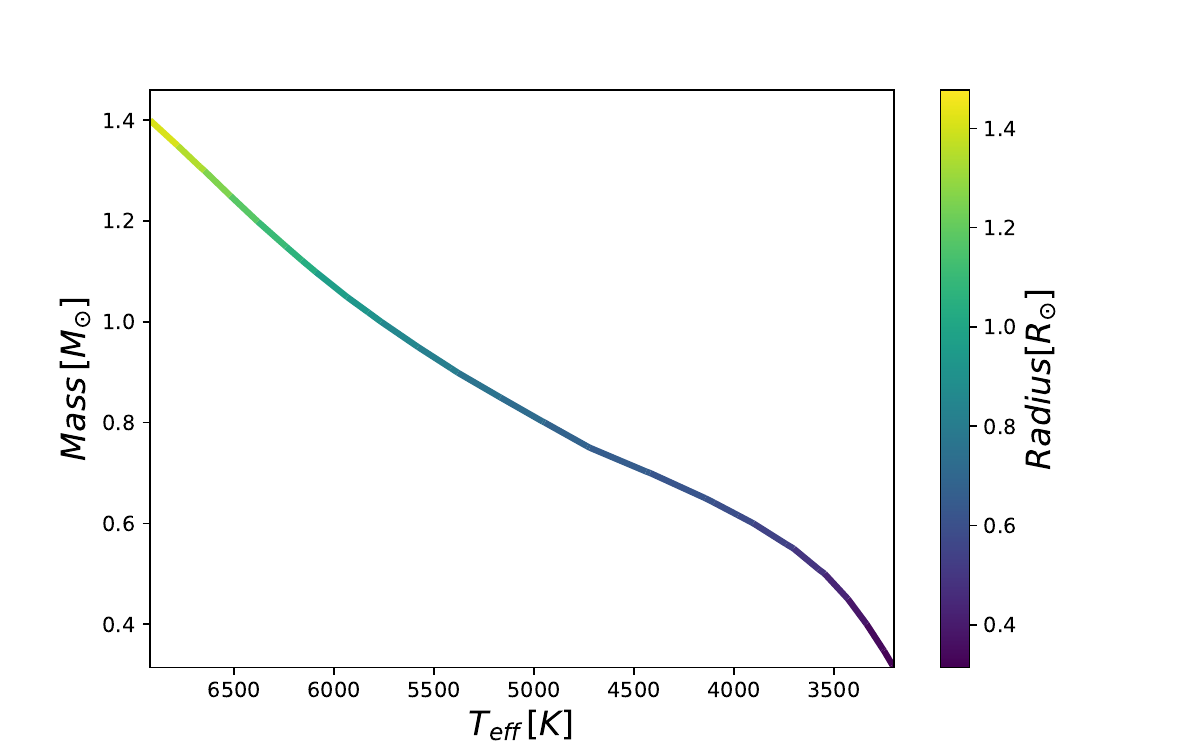}
    \caption{
    The correlation between the stellar masses and effective temperatures, color-coded by the radii of main-sequence stars from the isochrone. From left to right point the F, G, K, and M-type stars, respectively.}
    \label{fig6:central_star}
\end{figure}

This section aims to ascertain how the spectral types of low-mass host stars influence the radius valley. \S \ref{sec3.2.1} shows the FGKM-type host star data. \S \ref{sec3.2.2} primarily compares the impact of various spectral types of host stars on the mass loss and orbital migration of 
post-impact planets. In \S \ref{sec3.3}, we examine the impacts of 
the planetesimal impact on the radius distribution at 0.1 AU, 
ultimately confirming the potential formation of the radius valley through the planetesimal impact events. 
Furthermore, we also investigate the formation of a radius valley at the same irradiation energy distance in \S \ref{3.2.4}.
\subsubsection{Properties of Host Stars of Different Spectral Types }
\label{sec3.2.1}

Generally, FGKM-type host stars with masses ranging from 0.04-1.4 solar masses are primarily designated as main-sequence (MS) stars and classified based on their $\left(B-V \right)_{\rm 0}$ color bin. To get the stellar parameters of FGKM-type stars, we used the theoretical isochrone derived by the PARSEC models (version 1.2S) \citep{bressan2012parsec}\footnote{http://stev.oapd.inaf.it/cgi-bin/cmd}. Figure \ref{fig6:central_star} depicts the correlation between the stellar masses and effective temperatures, color-coded by the radii of MS stars aged 1 Gyr. 
Their masses and radii all increase with effective temperatures, which in turn influence the mass loss and orbital migration of post-impact planets.

Different spectral types of host stars have varying effects on the evolution of post-impact planets. These host stars play a significant role in determining the internal energy required for planetary mass loss. Specifically, the required core energy increases with decreasing host mass. We ignore the planetary evolution around the least massive M-type host stars to better understand the influence of the planetesimal impacts on planetary mass loss. 


\begin{figure}
    \centering
    \includegraphics[width=9cm]{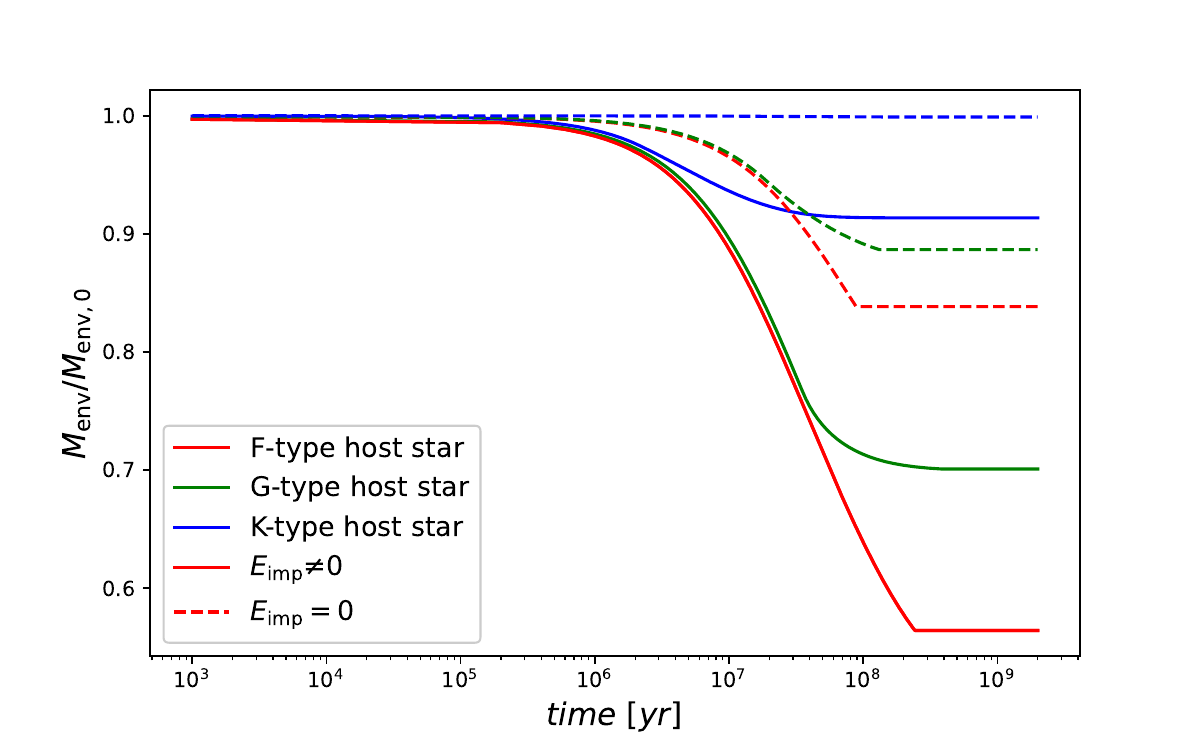}
    \caption{
    The effect of the planetesimal impacts on a planet's mass loss when orbiting F, G, and K-type stars.
    The colors red, green, and blue correspond to planets orbiting F, G, and K-type stars.
    The non-impact case is represented by the dashed line, while the giant-impact case is depicted by the solid line.
    The mass of the impactor is $M_{\rm imp} = 0.01 M{\rm c}$. The initial $\rm GCR$ is $0.03$ and the initial brightness temperature $T_{\rm b,0}$ is $5000K$.
    }
    \label{fig:central_star}
\end{figure}

\begin{figure}
    \centering
    \includegraphics[width=9cm]{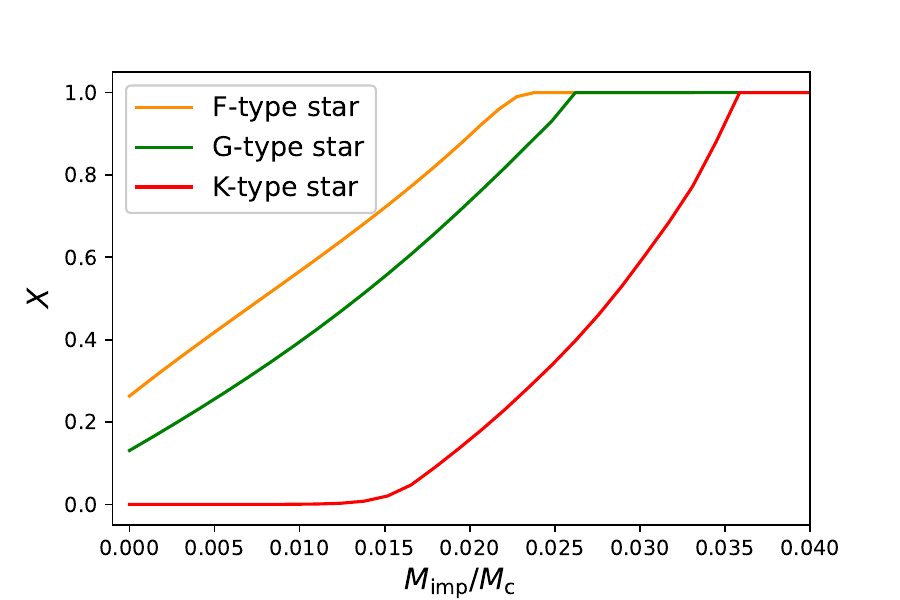}
    \includegraphics[width=9cm]{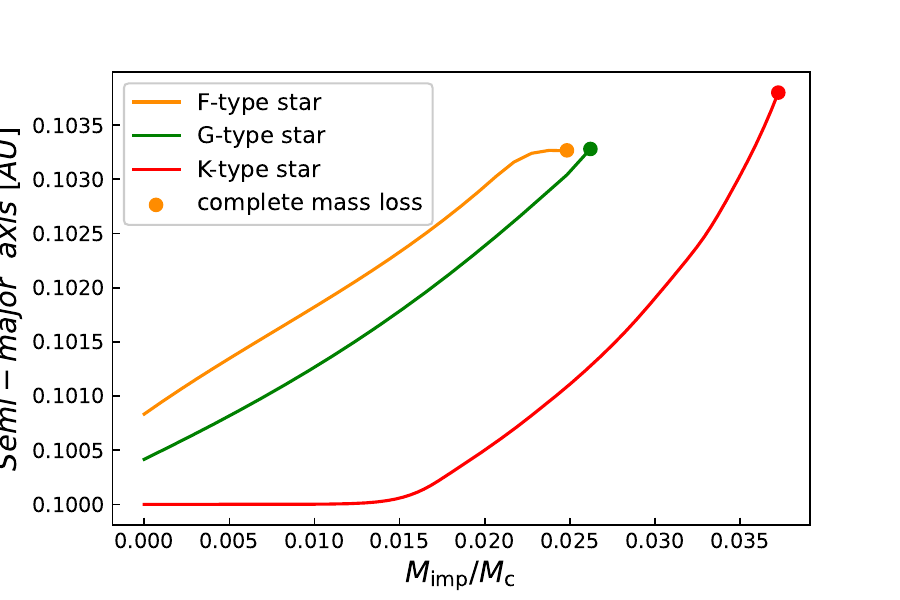}
    \caption{ The upper panel: the fraction of the H/He envelope lost $\left(X\right)$ of a planet orbiting FGK-type stars under the influence of various impactor masses.
    The lower panel: the relationship between the orbital semi-major axis and the impactor's mass.
    The orange, green, and red solid lines correspond to the scenario of planets orbiting FGK-type stars, respectively.
    The dot denotes the critical point for a complete mass loss.
    The initial conditions are identical to those depicted in Figure \ref{fig:central_star}.    
    }
    \label{fig4_different_star_X}
\end{figure}

\subsubsection{Effects of Spectral Types on Mass Loss}
\label{sec3.2.2}


This section mainly focuses on the mass loss 
of the post-impact planet around different spectral type stars, as shown by the solid lines in Figure \ref{fig:central_star}. Specifically, the blue, green, and yellow solid lines represent the post-impact planets around K, G, and F-type stars, respectively. The dotted lines represent the scenario without the collision.
The effective temperature and radius of the host star have a direct impact on $T_{\rm out}$, which is proportional to $T_{\star}R_{\star}^{1/2}$. As a result, an increase in both parameters from a K-type to an F-type star leads to a corresponding elevation of $T_{\rm out}$. 
Meanwhile, an increase in the mass of the host star results in a significant decrease in the planet's $R_{\rm out}$.
Based on the radial distributions of temperature and density in \S \ref{3.1.3_the_impact_mass}, the variation of the parameters on the RCB follows $T_{\rm rcb}\uparrow \,\, \longrightarrow\,\, R_{\rm rcb} \downarrow  \,\,\longrightarrow \,\, \rho_{\rm rcb} \uparrow $. The arrows indicate an upward direction for increasing values and a downward direction for decreasing values.
Furthermore, the increase in $c_{\rm s}$ due to its proportional relationship with $T_{\rm eq}^{\rm 1/2}$ is significant. Although $r_{\rm s}$ tends to decrease with the host star's mass and $c_{\rm s}$, its effect is too small to change mass loss. In addition, the significant increase in $c_{\rm s}$ can offset the impact of the reduction in $R_{\rm rcb}$. These parameters would results in a rise in $|\dot{M}_{\rm env}^{\rm B}|$.
Therefore, we posit that the mass-loss rate of the planet will exhibit a positive correlation with the host star's mass, resulting in an increased rate of orbital migration influenced by the mass-loss rate. Ultimately, due to the impact of both orbital migration and mass loss, planets revolving around the most massive F-type stars will eventually lose much mass.

Different types of host stars and collision energy are the primary factors influencing planetary mass loss in this section. In the absence of collisions, core energy drives this process. The extent of mass loss varies among different stars, shown in Figure \ref{fig:central_star}. As a host star transitions from an F-type to a K-type, its luminosity gradually reduces, resulting in reduced mass loss for the planet. However, if a planet experiences a collision, it leads to an increase in internal energy and consequently greater mass loss occurs. Therefore, impacted planets become lighter compared to unaffected ones.

 The mass-loss time satisfy that $\tau_{\rm loss} \sim M_{\rm env}/\dot{M}_{\rm env}$ (BS19) with the envelope mass $M_{\rm env}$ and the mass-loss rate of envelope $\dot{M}_{\rm env}$. 
Estimated mass-loss timescale ($\tau_{\rm loss}$) as a function of impactor mass ($M_{\rm imp}/M_{\rm c}$) for a range of H/He envelope mass fractions, which is shown in Figure 3 of BS19. Excluding the effect of {\bf the planetesimal} impact, all mass-loss timescales exceed 10 billion years. 
However, the collision heats the planet's core and envelope, causing a reduction in $\tau_{\rm loss}$ as $\dot{M}_{\rm env}$ accelerates (as seen in Figure \ref{fig:central_star}). Furthermore, $\tau_{\rm loss}$ of the impacted planet decreases with the increase in the host star's luminosity.
As the central star alters from K type to F type, a distinctive occurrence emerges: the time necessary for the planet to attain a stable mass state incrementally extends. This is attributed to the mass loss rate being dictated by  $\dot{M}_{\rm env}^{\rm E}$, not $ \dot{M}_{\rm env}^{\rm B}$.

To confirm the distribution of radii for post-impact planets around different spectral-type host stars, we need to consider the impactor's mass in relation to the mass loss and orbital evolution, as shown in Figure \ref{fig4_different_star_X}.  Specifically, the initial core-powered energies of planets around FG-type stars are higher than those around K-type stars actually. Figure \ref{fig4_different_star_X} shows that the initial $X$ in the upper panel increases with the host star's mass ($K \rightarrow F $ type), with planets around F-type stars showing the maximum value. Thus, the mass loss and orbital migration would increase significantly due to the impact. It should be noted that both $X$ and orbital migration increase with both host mass and impactor's mass, with both parameters reaching their respective maximum values upon the occurrence of a planet-impactor collision at critical mass, denoted as $M_{\rm{imp, crit}}$.

As illustrated by the solid red line in Figure \ref{fig4_different_star_X}, planets orbiting K-type stars experience a much decrease in mass and orbital displacement compared with other cases. However, if a sufficiently large impactor were to collide with a planet, it could completely lose its envelope and migrate beyond the orbital range for planets orbiting other host stars. Moreover, this orbital migration alters the planets' positions and increases their proximity to one another. This presents an intriguing signature for researching the variation in the radius distribution of planets occupying diverse orbital periods. 

\subsubsection{Radius Distribution }
\label{sec3.3}

\begin{figure}
    \includegraphics[width=9.5cm]{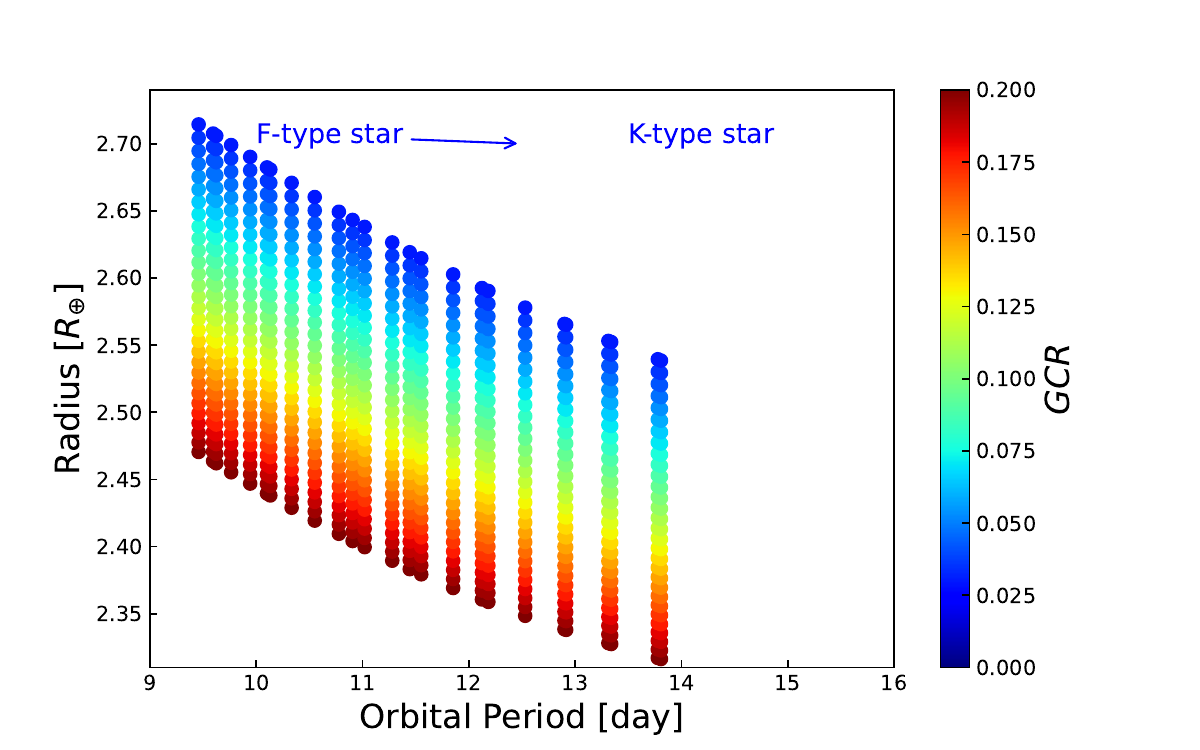}
    \includegraphics[width=9.5cm]{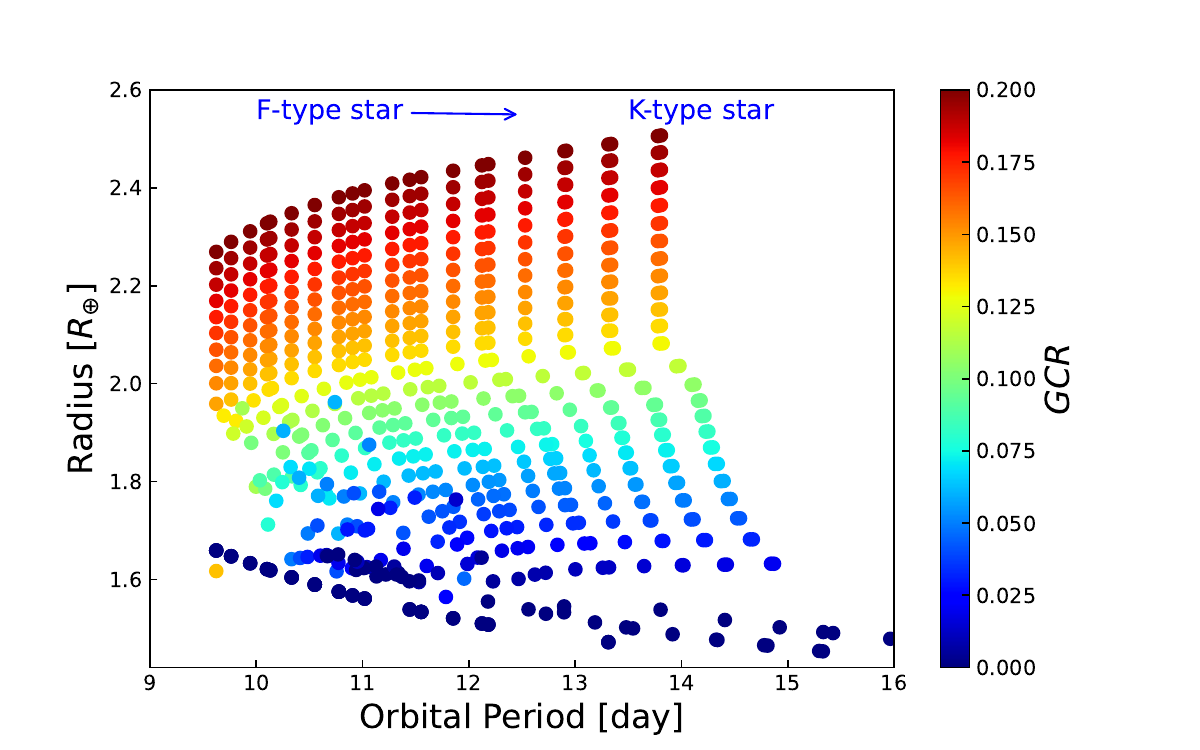}
    \caption{  Upper panel: The correlation between the initial orbit period and initial radius of the planet, color-coded by the final mass of the planetary envelope. Lower panel: The relationship between the orbit period and the radius of the post-impact planet after 2 Gyr.
    In this section, the impactor's mass is set to $M_{\rm imp}=0.05 \,M_{\rm c}$, and the initial orbital semi-major axis is $a = 0.1 AU$.
    }
    \label{fig5_radius}
\end{figure}

In this section, we focus on the effects of 
the planetesimal impact on planetary radii after 2 Gyr of evolution. 
The final radius (the photospheric radius) is determined by the RCB and the shallow radiative zone. It can be approximated as $R_{\rm p}\approx R_{\rm rcb}+\Delta R_{\rm atm}$, where $\Delta R_{\rm atm}$ represents the atmospheric correction \citep{2014ApJ...792....1L}.
The complete mass loss will occur when the $R_{\rm rcb}$ equals the core radius. Our calculations concentrate on studying the mass range of 0.03 to 0.2 times the core mass, which covers super-Earths and sub-Neptunes. Meanwhile, the core mass and radius of the planet remain the same as those adopted in \S \ref{sec3.2}. The planetesimal impact mechanisms rely on the initial core energy. Generally, the core energy is substantial during the magmatic ocean stage \citep{2018ApJ...Vazan}, resulting in a radius valley forming \citep{2018MNRAS.Ginzburg,2022MNRAS.Gupta}. However, to better understand the role played by the planetesimal impact mechanism, we will not consider the initial base temperature during the magmatic stage, instead setting it at 5000 K.

As depicted in the top panel of Figure \ref{fig5_radius}, the initial mass, radius, and orbital period distributions of pre-impact planets are presented. In this section, we assume that the initial semi-major axis of the planet is situated at 0.1AU. The orbital period of a planet is contingent on the host star mass, the planet mass, and the semi-major axis of the planet. Considering that the mass of the planet is significantly smaller than the mass of the host star, the initial orbital period distribution of planets revolving around the same host star remains consistent, as illustrated in the top panel. Upon collision, the planet experiences an instantaneous expansion and heating.

The lower panel of Figure \ref{fig5_radius} illustrates the correlation between a planet's radius and its orbital period after the planetesimal impacts. 
The final orbital period of a planet is determined by its semi-major axis and the mass of the host star it orbits. If a planet undergoes no orbital migration during its evolutionary process, both its orbital period and final mass will remain unchanged.  These planets would cool and contract.
However, the migration of a planet will result in an increase in its orbital period and additionally indicate the occurrence of mass loss.
The host star's mass has an inverse relationship with the orbital period. The orbital period increases from left to right in the lower panel of Figure \ref{fig5_radius}, and corresponding host stars' masses decrease. Therefore, a planet will have its maximum orbital period when orbiting around a massive host star.  Moreover, the initial masses of the planets in Figure \ref{fig5_radius}  consistently decrease from top to bottom.


\begin{figure}
    \includegraphics[width = 8cm]{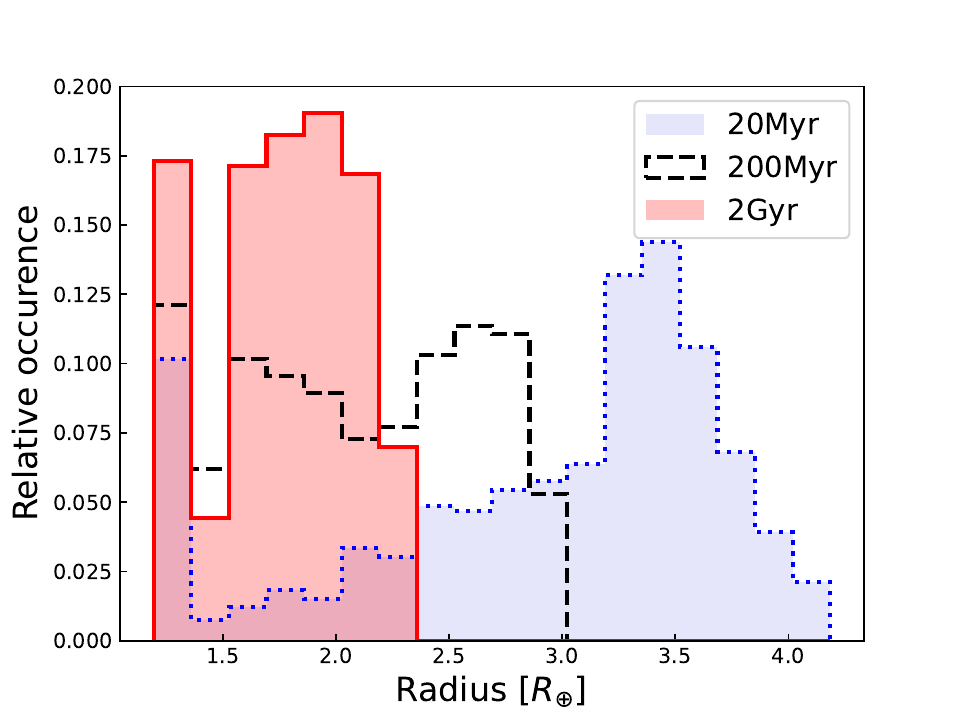}
    \includegraphics[width = 8cm]{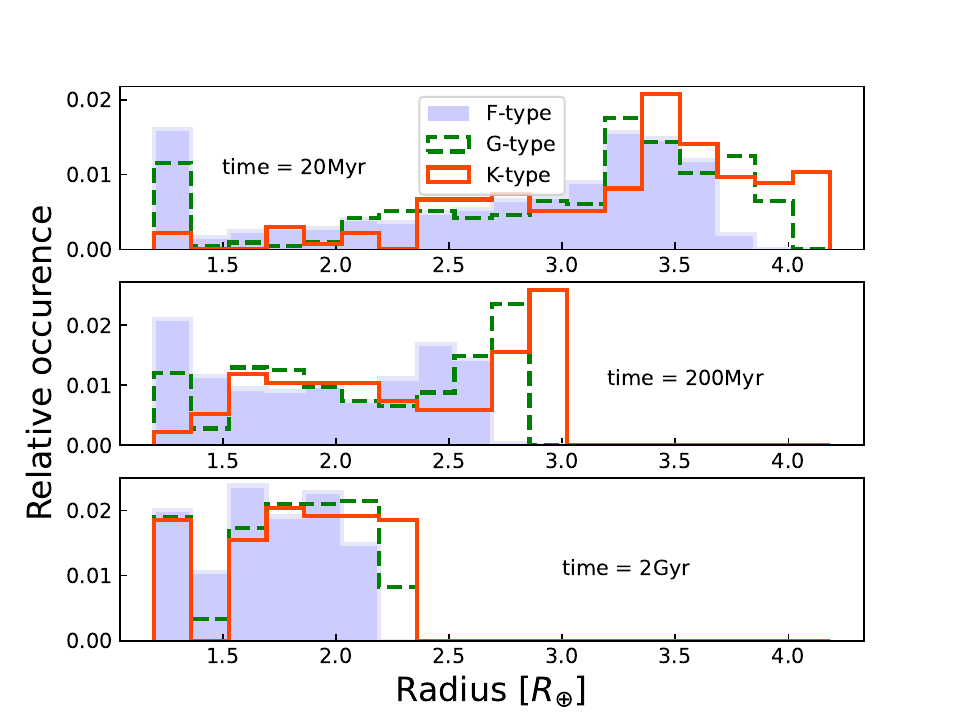}
    \caption{ The upper panel: the planetary radius distributions, showcasing the effects of evolution over 20 Myr, 200 Myr, and 2Gyr, from right to left. The lower panel: the variation in planetary radii around host stars of different spectral types over time, increasing from top to bottom. The purple area, green dotted area, and orange solid area correspond to the behavior of FGK-type stars respectively. It is important to note that the total planetary weight remains constant at 1.}
    \label{fig:radius_gap}
\end{figure}



There is a special radius of $\sim 1.85R_{\oplus}$ in the lower panel of Figure \ref{fig5_radius}.
For planets with a radius exceeding $1.85R_{\oplus}$, there is a direct positive correlation between their redness and their final evolved mass. The driving force behind this evolution is the increased core-powered energy by the planetesimal impact. Additionally, planets orbiting higher-mass host stars tend to cluster toward regions with smaller radii. As planets approach a radius below $1.85R_{\oplus}$, the increase in an orbital period of each column suggests that the planet's final mass diminishes and they take on a bluer hue. This is where the effect of the planetesimal impacts begins to emerge. Furthermore, as the 
 initial GCR of the planet decreases, the effects of the planetesimal impacts on the planet's evolution become more apparent.

 Smaller (the initial GCR is small) planets orbiting K-type stars exhibit a noticeable rightward shift in their orbital periods due to the effects of planetesimal impacts.  A radius gap near $1.6 R_{\oplus}$ exists between low-mass planets and those subjected to complete mass loss.  Additionally,  planetesimal impacts can cause significant core energy increases for planets orbiting hot F-type stars, making them highly susceptible to complete mass loss.  The final results deviate from the regularity of orbital period distribution observed around GK-type stars, which shows a wider distribution.  
Based on these results, a gap may be formed between these radii.

The radius distribution of post-impact planets is illustrated in Figure \ref{fig:radius_gap}, in which a gap is visible ranging from 1.3 $R_{\oplus}$ to 2.0 $R_{\oplus}$  after 20Myr (the light blue area), 200 Myr (black dotted line area) and  2Gyr (the red region).  At 20 Myr, there are peaks in the planet's radius distribution at 1.3 and 3.5 $R_{\oplus}$. However, both the more massive planets and the less massive ones will exhibit a reduced radius over time due to different processes: cooling and contraction for the former, and mass loss for the latter.  Eventually, the radius valley will form after 2Gyr.
This gap is believed to have resulted from the planetesimal impacts that are common in planets around FGK-type stars and is consistent with the radius valleys of $1.3-2.6R_{\oplus}$ observed in current Kepler planets. Although this gap differs from the $1.5-1.8R_{\oplus}$ radius valley by core-powered energy \citep{2018MNRAS.Ginzburg}, this discrepancy can be rectified by increasing the mass of impactors, increasing the mass of planets, and incorporating other isochronal data from K-type stars. Increasing the mass of the impactors can increase the efficiency of mass loss and migration, and the resulting impact on the overall radius will likely decrease. Additionally, the massive planets around many isochronal K-type stars can supplement existing data at a radius of $\sim 2.0R_{\oplus}$. By comprehensively adjusting these influences, the ultimate radius valley will likely align with the $1.5-1.8 R_{\oplus}$ range.


In conclusion, the planetesimal impacts, especially those accompanied by orbital migration, have been demonstrated to effectively contribute to the mass loss of planets located at a distance of 0.1 AU.
By carefully selecting the masses of planets and impactors, as well as their host stars, we can create a distinct radius valley for planets in prominent positions. This can be achieved by distinguishing between smaller super-Earths and larger sub-Neptunes. Therefore, we believe that a properly parameterized planetesimal impact can facilitate the formation of such a radius valley.

\subsubsection{The Formation of Radius Valley at Same Irradiation Energy Distance}
\label{3.2.4}

\begin{figure}
    \centering
    \includegraphics[width = 9 cm]{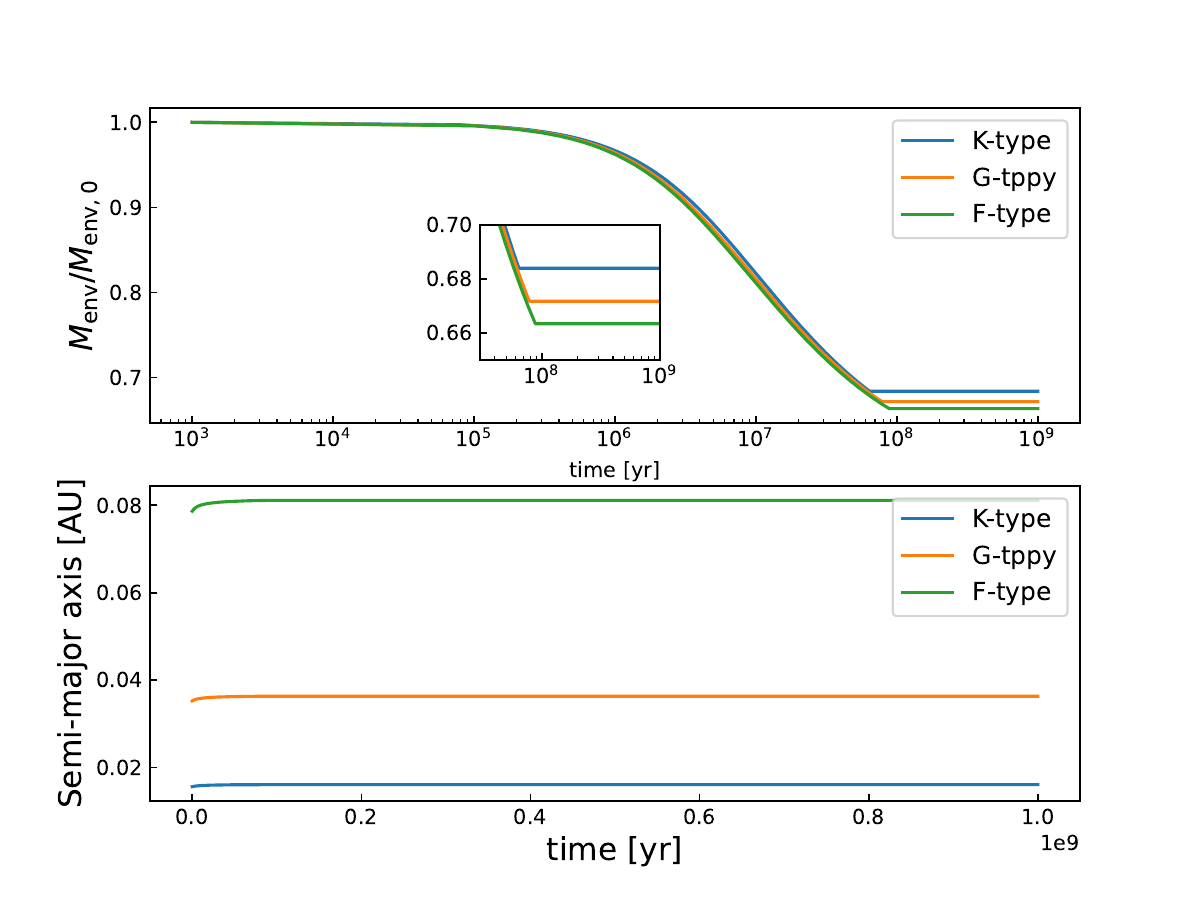}
    \caption{ The upper panel shows the mass loss of planets orbiting different host stars at the same irradiation energy distance and the lower panel corresponds to the evolution of the semi-major axis of their orbits. The solid blue, orange, and green lines correspond to planets orbiting K, G, and F types, respectively. The settings of other parameters in the calculation are consistent with those in Figure 6.}
    \label{fig:same_irradiative_flux_distance}
\end{figure}

 Previous calculations have assumed a constant semi-major axis of the initial orbit as the initial distance between the host star and the planet in different spectra. However, the choices of initial conditions in this paper are not reasonable. Planets will migrate within their natal protoplanetary disc, positioning themselves at the inner edge of the disc if they move towards the center. Typically located at the silicate sublimation front, which is determined by each star and maintains a constant temperature, this is where they will begin to receive irradiation later on.

 Stellar luminosity determines the position of the sublimation front \citep{2010ARA&A..Dullemond}. The sublimation temperature follows  $T_{\mathrm{s}}=2000 \left({\rho}/{1 \mathrm{~g} \mathrm{~cm}^{-3}}\right)^{0.0195} \ \mathrm{~K}$ \citep{2005A&A...Isella,2019A&A...Flock} with the density at silicate sublimation front. In general, the sublimation temperatures for $Mg_{2}Si0_4$ and $MgSi0_3$ are $1354 {\rm K}$ and $1500 {\rm K}$, respectively. For this study, we will assume a sublimation temperature of $1500 K$.  The initial semi-major radius of the planets orbiting the host star with different spectral types, located in front of silicate sublimation, can be determined by the equilibrium temperature resulting from the planet's interaction with the star. Consequently, the initial semi-major axis determines the outer boundary radius of the planet, potentially altering its mass loss process.

When the Bondi radius determines the outer boundary of a planet, planets orbiting host stars of different spectral types will have identical radii. Consequently, planets located at the same irradiation flux distance will exhibit identical mass loss histories. However, when the Hill radius determines the outer radius, there will be variations in the tendency of mass loss.  
As depicted in the lower panel of  Figure \ref{fig:same_irradiative_flux_distance}, when the central star transitions from a K type to an F type at the same radiative energy distance, both the initial semi-major axis and Hill radius of the planet will increase.
The increase in radius results in a slight elevation of mass loss due to the decrease in gravitational energy ($\left| E_{\rm g}\right| \approx G M^2/R_{\rm out}$, \citealt{2013sse..book.....Kippenhahn}) at RCBs for a planet with equal mass. 
As depicted in the upper panel of  Figure \ref{fig:same_irradiative_flux_distance}, F-type planets demonstrate a slightly greater capacity for mass loss. 
Therefore, we can conclude that as the host star changes from K-type to F-type at identical irradiation flux distances, there is a gradual increase in the planet's ability to lose mass.

\begin{figure}
    \includegraphics[width = 8cm]{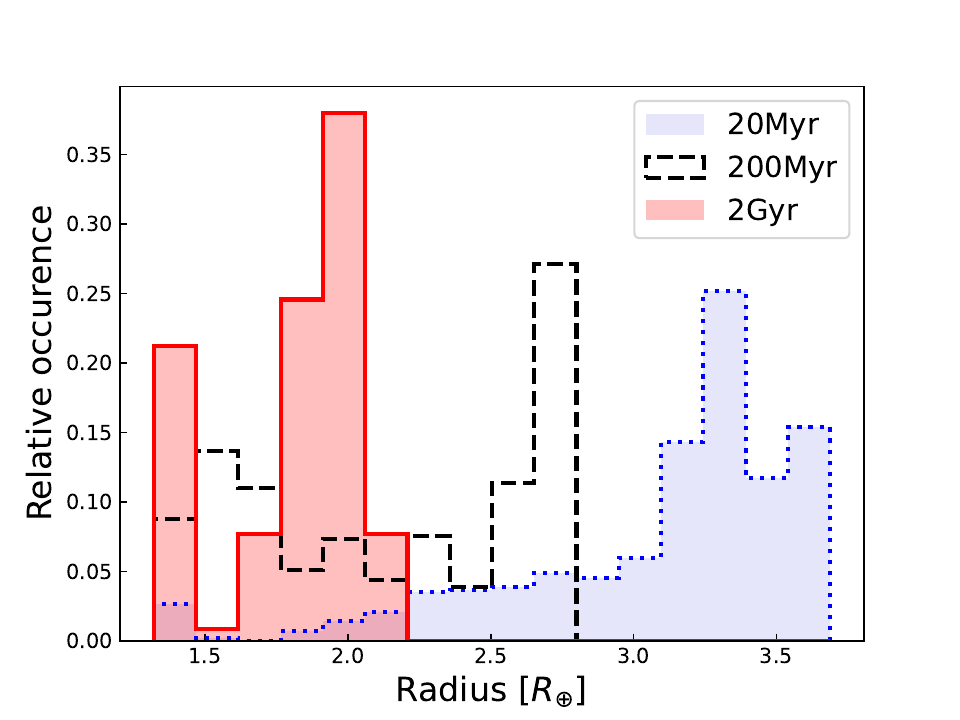}
    \includegraphics[width = 8cm]{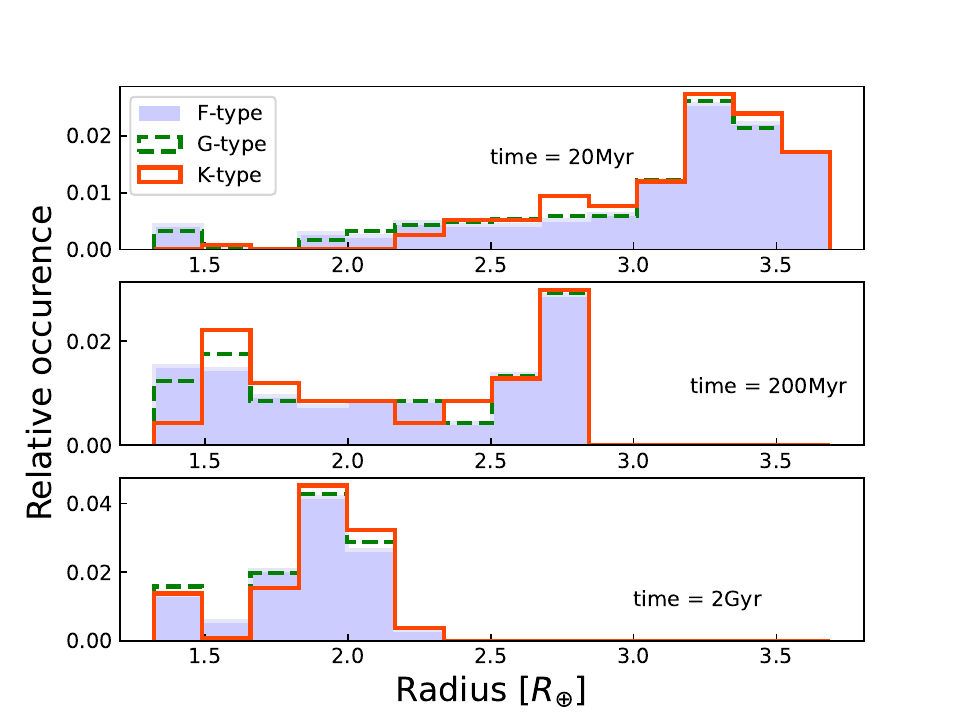}
    \caption{ The upper panel illustrates the distribution of planetary radii across different evolutionary timescales. The light blue area represents a timeframe of 20 Myr, the black dashed area represents 200 Myr, and the light red area indicates 2 Gyr. On the lower panel, we can observe how the radius distribution of planets is influenced by various spectral types of host stars at different stages of evolution. Specifically, F-type host stars correspond to the purple area, G-type host stars correspond to the green dashed area, and K-type host stars correspond to the red solid area.}
    \label{fig10}
\end{figure}

This section examines the relationship between a planet's initial semi-major axis and the same irradiation energy distance, which ultimately determines its evolutionary outcomes following impacts.  Figure \ref{fig10} illustrates that at 20 Myr, there is a concentration of larger radii in the distribution of the post-impact planets.  As evolution progresses, smaller peaks emerge on the left side and eventually form a valley of $1.3-2.0R_{\oplus}$.  The formation of this valley is influenced by FGK-type host stars, as depicted in Figure \ref{fig:same_irradiative_flux_distance}.  Planets orbiting F-type stars experience significant mass loss after 20 Myr. However, planets orbiting different types of host stars exhibit little variation in their final mass when exposed to the same irradiation energy distance. Consequently, planets exhibit minimal variation in radii after two billion years on the left side of Figure 3.  Nevertheless, planets around the F-type host star complement the data in the lower region of the valley and will contribute to the formation of super-Earth/sub-Neptune in that specific range.


Our primary focus is studying the atmospheric mass loss of planets during their magma ocean phase after impact events around FGK-type stars.  However, in the Praesepe cluster detected by NASA's K2 mission,  a significantly high incidence rate ranging from 79\% to 107\% for young and hot sub-Neptus around these stars \citep{2023AJ..Christiansen}.  This occurrence rate in both Praesepe and Hyades clusters is consistent with that predicted based on the core-driven mass loss mechanism \citep{2023AJ..Christiansen}.  Therefore, future studies can also investigate the mass loss of GKM-type stars after impacts from hot sub-Neptune.


\section{Conclusion and discussion}
\label{section_conclusion}

The planetesimal impact could cause planetary mass loss, resulting in outward migration. This outward migration can suppress further mass loss and ultimately impact the distribution of planet radii.   Moreover, the post-impact planet's mass loss could also be significantly affected by the type of star in their star-planet systems.    Therefore, it is necessary to investigate whether different spectral types of stars contribute to different radius distributions and could cause a radius valley to emerge.

The final radius of a planet is determined by how it loses mass and moves in its orbit, especially around different types of stars.    Larger and heavier host stars can heat up the outer layers of planets, leading to more mass loss and further migration.     Consequently, planets in close proximity to massive F-type stars experience substantial mass loss and undergo significant migration.
Obvious gaps in the final radius are observed between planets characterized by small and massive initial GCR. 
However, planets around K-type stars show a moderate mass loss, where orbital migration suppresses the mass loss and then maintains the final radius at a higher level.    This signature would create a gap near $1.6 R_{\oplus}$ before the complete mass loss.    Combined with these effects, a valley of radius between {\bf $1.3 -2.0 R_{\oplus}$} would be formed, which is consistent with the observed valley of Keplerian radius between $1.3-2.6 R_{\oplus}$.      From this, we can conclude that the planetesimal impacts may promote the radius valley forming. 

There is a slight discrepancy in the radius valley between $1.5-1.8R_{\oplus}$ driven by core-powered energy \citep{2018MNRAS.Ginzburg,2020MNRAS.493..Gupta}, 
but we can employ several methods to reconcile our results with observations in the future. 
A planet's base atmosphere can be heated by the planetesimal impact, even if the core energy is less massive. This is because the impact energy would increase the core energy.
In this work, we will exclude the change in the planet's radius caused by the huge core-powered energy during the magma ocean stage. Instead, we concentrate on the moderate initial core energy for investigating the role of the planetesimal impact.
Several factors contribute to the uncertainty of the planetesimal/giant impact process, including the number of collisions, the impactor's mass, and planetary mass. Firstly, during the dynamic instability phase, most planets experience one or two planetesimal/giant impacts with only a few experiencing up to four \citep[N-body collision]{2022ApJ...Izidoro}. However, the appropriate number of the planetesimal/giant impacts can promote the formation of the radius valley. 
Secondly, if a planet has low core energy or greater initial GCR, increasing the impactor's mass can significantly enhance its core power or even result in complete mass loss and subsequent migration. Consequently, massive impactors will form numerous super-Earths around $1.5R_{\oplus}$ whose envelopes have been stripped off at the first peak of radius distribution \citep{2020A&A...Venturini}. Without other mechanisms involved, a second peak at $1.8R_{\oplus}$ may also be formed when complete mass loss does not occur. Additionally, Mini-Neptunes may contain high-mean molecular weight compounds and $H_{2}O$ \citep{2021MNRAS.Biersteker}. These substances could be lost due to photoevaporation or other processes \citep{2020A&A...Venturini,2022ApJ...Izidoro}, leading to their concentration on the second peak.  This method of naturally forming a radius valley may be supported by observational data.  In conclusion, although the exact parameters are uncertain, we believed that the planetesimal/giant impact with specific conditions plays a crucial role in promoting the formation of a radius valley for the planets around FGKM-type host stars.




The main focus of our study is to investigate the impact of planetesimals on mass loss and the attempt to reproduce the radius gap. We have found that there are significantly more planetesimal impacts than actual giant impacts, resulting in a much greater mass loss. However, it remains questionable whether reproducing the radius gap can be achieved with only a few planetesimal impacts. In future studies, we plan to provide further evidence by conducting additional tests specifically targeting planetesimal impacts.

Some caveats of our results are supposed to be remedied for further investigations.   First, temperature fluctuations in the radiative layer were overlooked.   
The planetary atmosphere model incorporates irradiated atmospheres \cite{2010A&A...520A..Guillot,2015A&A...574A..Parmentier}, which involves radiative transfer. The presence of strong absorption at visible wavelengths leads to significant temperature inversions in the uppermost layers. Introducing radiative energy as a new energy input may elevate temperatures on the RCB and contribute to mass loss, although this impact might not be explicitly visible in the thinner radiative layer.
Second, the planetary orbital motion during a major collision was not considered. 
If this is included in the analysis of planetary mass loss, it becomes necessary to address the issue of three-body dynamics and provide a comprehensive explanation of the planet's orbital evolution \citep{2019Natur.Liu}. 
Moreover, the interrelation between orbital migration during the impact and post-impact stages will significantly modify the overall mass loss process.  
The orbital migration and mass loss resulting from photo-evaporation can alter the dynamical evolution of multiple terrestrial planetary systems \citep{2023AJ....165..174W}. This mechanism may also apply to the mass loss induced by giant impacts.
This research explores the radius valley by utilizing a single fixed semi-major axis and core mass. To yield more authentic outcomes, the orbital period and core mass distribution proposed by \cite{2017ApJ...847...Owen} could be incorporated in future studies.
Thus, if the aforementioned factors are considered, the interaction of mass loss and planetary orbital evolution would be better constrained.

Recent studies on planetary mass loss 
focus on 
photo-evaporation \citep{2017MNRAS.Owen} and core-powered mass loss \citep{2018MNRAS.Ginzburg,2022MNRAS.Gupta}. 
Planesimal impacts can change the ice-rock ratio on the planet, resulting in a planet made of water-ice and rock around its star \citep{2023MNRAS.Lozovsky}. Moreover, giant impacts can turn a planet with a lot of volatiles into a planet with less volatiles \citep{2015ApJ...Liu}. 
\cite{2014ApJ...Schlichting} suggested that the formation of close-in super-Earths and mini-Neptunes was likely significantly influenced by the migration of either solid materials or fully formed planets.
The period vs. radius distribution of planets can be recreated by evaporation and giant impact, as long as the initial radial distribution of the core mass follows a wide range of power-law distributions and the mass fraction of the envelope is around 0.1 \citep{2021ApJ...Matsumoto}.
Even in the absence of these mechanisms, planets lose mass when their protoplanetary disks disperse.
This mass loss can spontaneously induce the radius valley \citep{2022ApJ..Lee}.
In conclusion, the planetesimal/giant impact resulted in significant mass loss, which played a crucial role in shaping the formation and internal composition of near-Earth super-Earths and sub-Neptunes.
By comparing various mass loss mechanisms, further physical insights can be obtained about the formation process of super-Earths.


\section{acknowledgments}
We thank the anonymous referee for his/her suggestions that greatly improved this paper.
This work has been supported by the National SKA Program of China (Grant No. 2022SKA0120101) and National Key R \& D Program of China (No. 2020YFC2201200) and the science research grants from the China Manned Space Project (No. CMS-CSST-2021-B09 \& CMS-CSST-2021-A10) and the grants from The Macau Science and Technology Development Fund, Macau SAR (File No. 0001/2019/A1, No. 0051/2021/A1) and the National Natural Science Foundation of China (grants 42250102), and opening fund of State Key Laboratory of Lunar and Planetary Sciences (Macau University of Science and Technology) (Macau FDCT Grant No. SKL-LPS(MUST)-2021-2023). C.Y. has been supported by the National Natural Science Foundation of China (grants 11521303, 11733010, 11873103, and 12373071).






\appendix
\section{The Evolution of Post-impact Planets}
\label{method}

In order to facilitate the calculation of planetary evolution, we make the assumption of a constant radial luminosity denoted as $L = L_{\rm rcb}$, which in turn simplifies the energy conservation equation (\ref{eq4:energy_conservation}) within the equation for the internal structure of the planet. Additionally, \cite{2014ApJ...Piso} proposed that the temperature of the radiative layer is approximately equal to the outer temperature, that is $T \approx T_{\rm out}$, resulting in the elimination of Equation (\ref{eq3:energy_transfer}) and the radiative temperature gradient within the radiative layer. However, in order to calculate the temperature variation in the convective layer, it is imperative to employ the energy transport Equation (\ref{eq3:energy_transfer}) while substituting $\nabla$ with $\nabla_{\rm ad}$.
By combining the newly derived equations of planetary structure and energy, we can analyze a planet's planetary structure and evolution.

The initial model entails the determination of the atmospheric envelope's parameters by assuming the gas-to-core ratio, referred to as GCR. This ratio serves as the foundation for the subsequent evolution analysis of the planet. The mass of the envelope at the subsequent time step is denoted as 
\begin{equation}
M_{\rm n e x t}=M_{\rm e n v, pre}+\dot{M}_{\rm e n v} \Delta t,
\end{equation}
where $M_{\rm e n v, pre}$ is the  envelope mass at previous time, and $\Delta t$ is the timescale between two adjacent snapshot.
However, the ensuing energy is obtained as
\begin{equation}
E_{\rm n e x t}=E_{\rm c0}+E_{\rm t h0}+E_{\rm g0}+\left(\dot{E}_{\rm e n v, m0}+\dot{E}_{\rm e n v, L0}\right) \Delta t, 
\label{eq:next_energy}
\end{equation}
where the symbols $E_{\rm c0}$, $E_{\rm t h0}$, $E_{\rm g0}$, $\dot{E}_{\rm e n v, m0}$ and $\dot{E}_{\rm e n v, L0}$ are the core energy, thermal energy, gravitational energy, energy change of mass loss and luminosity at the previous time, respectively. When accounting for the planet's self-gravity in the process of mass loss, the crucial aspect of planetary evolution lies in the resolution of the corresponding base temperature ($T_{\rm b}$) and base density  ($\rho_{\rm b}$), given the known planet mass and total energy. We use Newton's iteration to solve the following system of equations to obtain the planetary atmospheric base parameters of the atmosphere, i.e.,
\begin{equation}
f_1\left(T_b, \rho_b\right)=E\left(T_b, \rho_b\right)-E_{\text {next }}=0,
\end{equation}
\begin{equation}
f_2\left(T_b, \rho_b\right)=M\left(T_b, \rho_b\right)-M_{\text {next }}=0,
\end{equation}
where the symbols $E_{\text {next }}$ and $M_{\text {next }}$ correspond to the energy and mass at the next time, respectively. Once the planet is collided by an object with an appropriate mass,  in which the impact energy $E_{\rm imp}$ would be induced in the planet. The total energy in Equation (\ref{eq:next_energy}) will be increased. Consequently, the alteration of the planet's core temperature and evolution would ensue. 

\section{The effect of self-gravity}
\label{sec:self_gravity}

The stability of a planet's atmosphere is determined by its self-gravity, which plays a crucial role. As the planetary mass increases, so does the self-gravity. A planet with greater mass (stronger self-gravity) requires much more energy to escape the atmosphere. We will examine the effect of the self-gravity on atmospheric loss.

Excluding the planet's self-gravity, its total mass is approximately equal to its core mass, i.e., $M_{\rm p}$ should be replaced with $M_{\rm c}$. Hence, the equation of mass conservation will be eliminated, and in other equations of planetary structure, the notation $M_{\rm r}$ representing the radial mass at any shell $r$ should be replaced with the core mass $M_{\rm c}$. The envelope mass (BS19) becomes
\begin{equation}
M_{\mathrm{env}} 
= 4 \pi R_{\mathrm{c}}^3 \rho_{\mathrm{b}} \int_1^{x_{\mathrm{rcb}}}\left[\nabla_{\mathrm{ad}} \Lambda\left(\frac{1}{x}-1\right)+1\right]^{\frac{1}{\gamma-1}} x^2 \mathrm{~d} x \ ,
\end{equation}
where the symbols $x$ and $x_{\rm rcb}$ define as $x=r / R_{\mathrm{c}}, x_{\mathrm{rcb}}=R_{\mathrm{rcb}} / R_{\mathrm{c}}$, respectively. In addition, $\Lambda \equiv G M_{\mathrm{c}} \mu /\left(R_{\mathrm{c}} k_{\mathrm{B}} T_{\mathrm{b}}\right)$. $\gamma$ is the adiabatic index of the envelope.
The gravitational energy and the  thermal energy (BS19) should become as follows
\begin{equation}
E_{\mathrm{g}} 
 = -4 \pi G M_{\mathrm{c}} \rho_{\mathrm{b}} R_{\mathrm{c}}^2 \int_1^{x_{\mathrm{rcb}}}\left[\nabla_{\mathrm{ad}} \Lambda\left(\frac{1}{x}-1\right)+1\right]^{\frac{1}{\gamma-1}} x \mathrm{~d} x \ , 
\end{equation}
\begin{equation}
 E_{\mathrm{th}} 
 = 4 \pi \rho_{\mathrm{b}} R_{\mathrm{c}}^3 \frac{k_{\mathrm{B}} T_{\mathrm{b}}}{\mu(\gamma-1)} \int_1^{x_{\mathrm{rcb}}}\left[\nabla_{\mathrm{ad}} \Lambda\left(\frac{1}{x}-1\right)+1\right]^{\frac{\gamma}{\gamma-1}} x^2 \mathrm{~d} x \ .   
\end{equation}
BS19 also present the results of the post-impact planet in a non-self-gravity case.

\begin{figure}
    \includegraphics[width = 9cm]{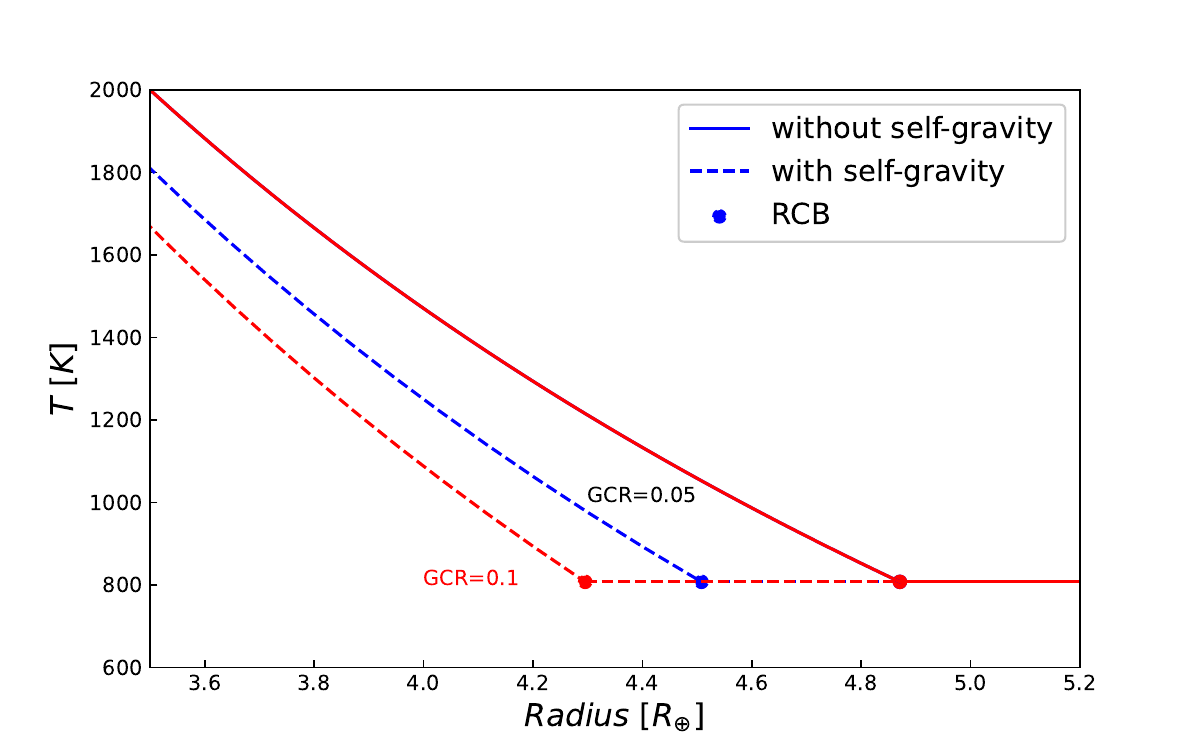}
    \includegraphics[width = 9cm]{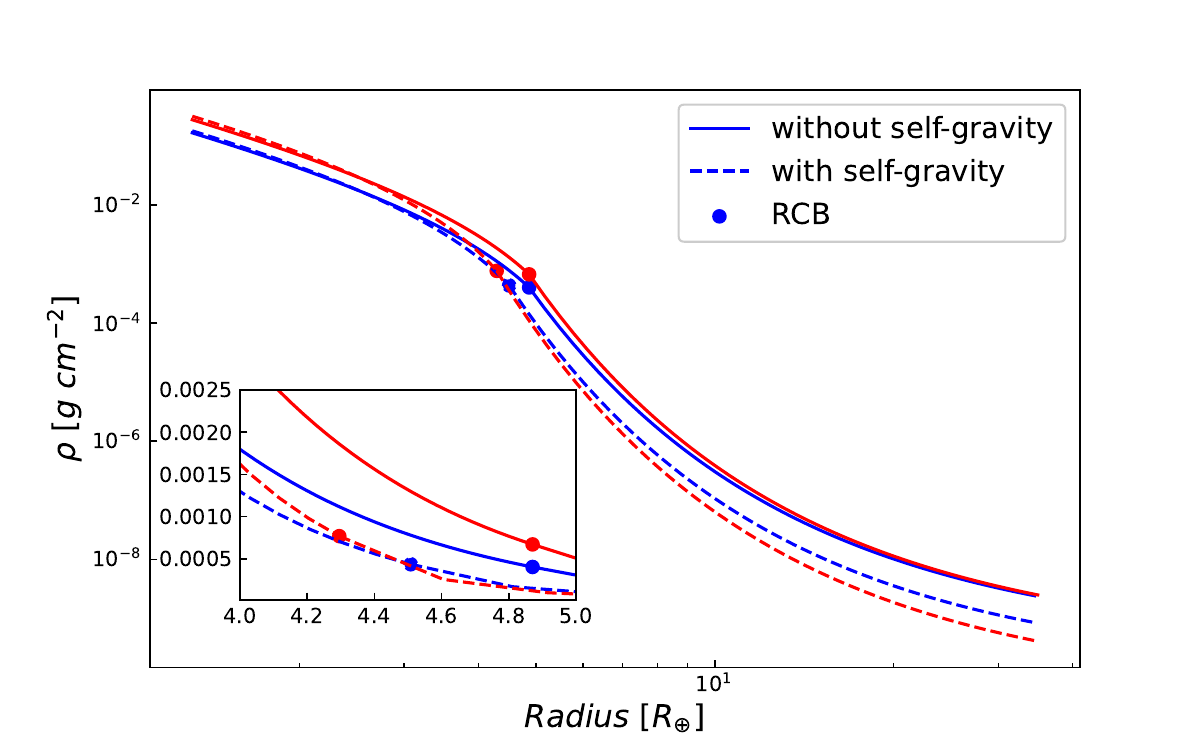}
    \caption{ The panels on both sides display different aspects: one shows radial temperature while the other illustrates radial density. The black lines are utilized to indicate a gas-to-core mass ratio of 0.5, whereas a red line signifies a ratio of 0.1. Without considering self-gravitation, the radial temperature profiles of both cases align perfectly. The radiative-convective boundary represented by the circle dot separates the convective zone on the left side from the radiative zone on the right side. In order to distinguish between scenarios with or without self-gravity effects, solid and dashed lines are employed respectively.}
    \label{fig:radial_structure}
\end{figure}

\begin{figure}
     \includegraphics[width=9cm]{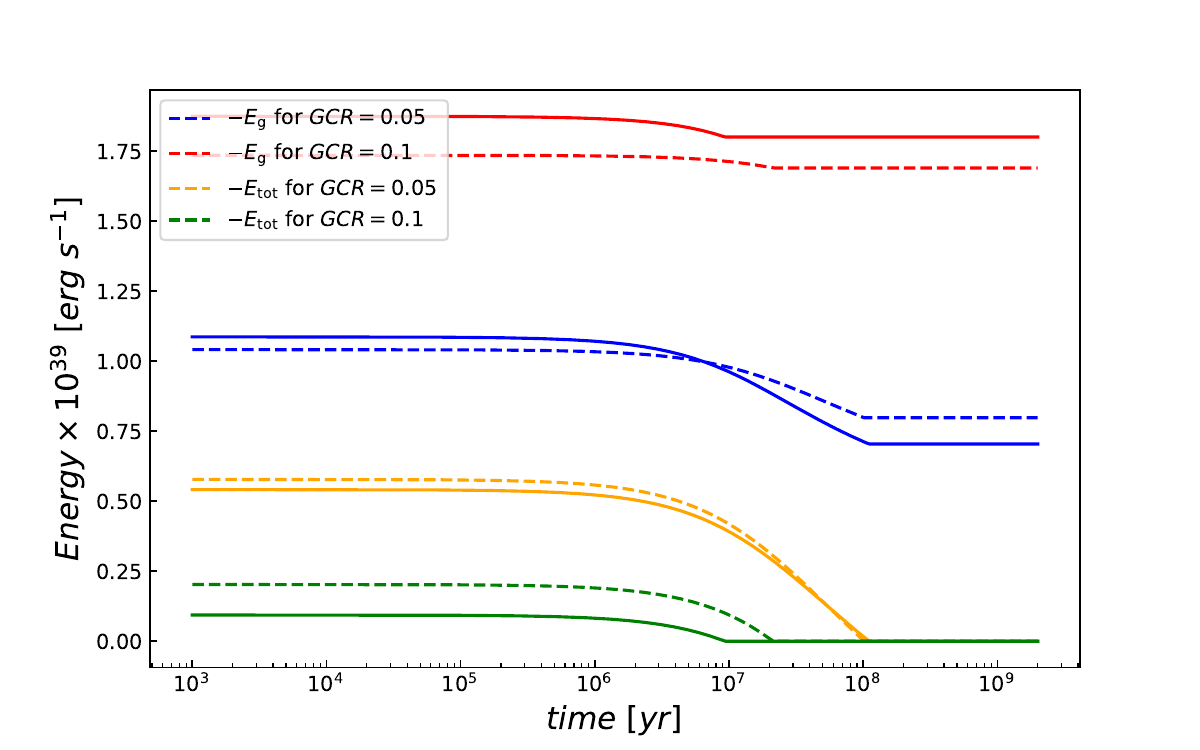}
    \includegraphics[width=9cm]{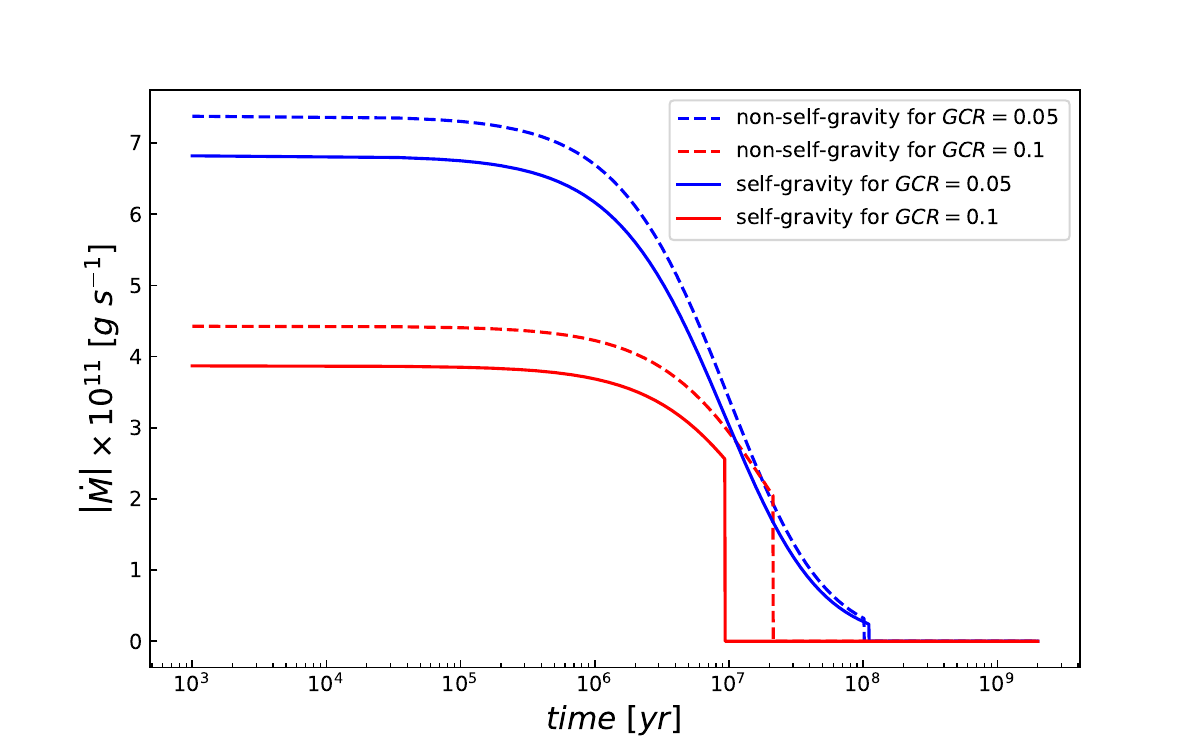}
    \caption{ The left side of the figure depicts the variations in gravitational energy and total energy in a mass loss model with initial gas-to-core mass ratios of 0.05 and 0.1, respectively. Solid lines represent the presence of self-gravity, while dotted lines indicate its absence. The blue and red lines correspond to the gravitational energy for GCR values of 0.05 and 0.1, respectively, while the orange and green lines represent the total energy for GCR values of 0.05 and 0.1, respectively. On the right side, the corresponding mass-loss rate is shown with blue indicating the case for GCR = 0.05 and red representing the case for GCR = 0.1.}
    \label{fig:enter-label}
\end{figure}

To examine the impact of self-gravity, we assign initial GCR values of 0.05 and 0.1, respectively. The planet's initial base temperature is set at 9000K, and its orbit has a semi-major axis of 0.1 AU. The cases presented by BS19 do not consider self-gravity, whereas we will examine the cases that take self-gravity into account. Figure \ref{fig:radial_structure} illustrates the initial radial structure of the planet's model.

Firstly, the left panel's radial temperature diagram shows that self-gravity (represented by the blue dashed line) can push the planet's radiative convection boundary (i.e., RCB)  to move inward. This indicates a decrease in radiative energy flux density within the convection region, resulting in an increased mass loss time ($\tau_{\rm loss}$) and the final mass of the planet. The right panel's radial density relation also demonstrates that self-gravity leads to a lower density at the outer boundary compared to the non-self-gravity case. Consequently, under identical orbital radius conditions, there will be a decrease in mass-loss rate and extended mass-loss time ($\tau_{\rm loss} \sim M_{\rm env}/  \dot{M}_{\rm env}$) similar to what was observed in temperature analysis.

Secondly, we predict that as the initial mass of the planet increases, the effect of self-gravity will also increase. As shown on the left side of this figure, when self-gravity is absent, both the radial temperature profile and position of RCB remain unchanged regardless of how much we increase its initial total mass. Only when the radial density profile experiences a reduction, specifically at the RCB, causing a drop in density at the outer boundaries, does it result in a decline in mass loss rate and an extension of the mass-loss time. When considering self-gravity effects, this enhancement expands such phenomena observed from ultimately leading to an extension of mass-loss time and allowing for greater retention of planetary mass.

Figure \ref{fig:enter-label} illustrates the changes in energy and mass-loss rate for both cases during the mass-loss process mentioned above. The left panel demonstrates the impact of self-gravity on parameters, represented by the solid curve, compared to its absence depicted by the dotted curve.  This change results in a varying radial mass instead of being approximately equal to the planet's core mass.  This realization leads to an increase in gravitational energy within the planet. However, the gravitational potential energy of both the self-gravity and non-self-gravity cases for GCR = 0.06 can intersect over time.  
This occurs primarily because of a more significant decline in core temperature within systems influenced by self-gravity. However, the total energy levels 
are consistently lower for systems governed by self-gravitational forces compared to those without self-gravity cases. 

On the left panel, a planet's self-gravity significantly reduces its rate of mass loss and enhances its ability to retain more mass compared to when it lacks self-gravity.  A planet can directly reduce its rate of mass loss to zero once its total energy reaches zero since the mass-loss rate is determined by $\dot{M}_{\rm env}^{\rm B}$ instead.  When considering scenarios with low self-gravity, there is no significant difference in the timescale required for achieving zero mass loss.   However, as a planet's self-gravity increases along with the corresponding increase in GCR, $\dot{M}$ takes less time to reach zero value.  
This signature suggests that planets with stronger gravitational forces stop losing mass at an earlier stage ($t_{\rm loss} \approx M_{\rm env}/\dot{M}$) and preserve larger portions of the atmospheres through a cooling contraction process. 
The initial mass range we have selected for our model is between 0.05 and $0.2M_{\rm c}$. When dealing with larger planets, the energy required for complete mass loss is greater. The rate at which a planet loses mass during an impact is limited by the mass and escape velocity of the impacting object. Therefore, achieving complete mass loss on a larger planet would necessitate an even larger impacting object.

\bibliography{sample631}{}
\bibliographystyle{aasjournal}



\end{document}